\documentclass[aip, amsmath,amssymb, preprint]{revtex4-1}

\usepackage[geometry]{ifsym}
\usepackage{pstricks}
\usepackage{graphicx}
\usepackage{dcolumn}
\usepackage{bm}

\usepackage[utf8]{inputenc}
\usepackage[T1]{fontenc}
\usepackage{mathptmx}
\usepackage{etoolbox}

\makeatletter
\def\@email#1#2{%
 \endgroup
 \patchcmd{\titleblock@produce}
  {\frontmatter@RRAPformat}
  {\frontmatter@RRAPformat{\produce@RRAP{*#1\href{mailto:#2}{#2}}}\frontmatter@RRAPformat}
  {}{}
}%
\makeatother
\begin{document}

\preprint{AIP/123-QED}

\title{Existence of a maximum flow rate in electro-osmotic systems} 

\author{Sleeba Varghese}
\affiliation{ 
Sorbonne Université, CNRS, Physico-chimie des Électrolytes et Nanosystèmes Interfaciaux, PHENIX, F-75005 Paris}%
\author{Billy D. Todd}
\affiliation{Department of Mathematics, School of Science, Computing and Engineering Technologies, Swinburne University of Technology, Melbourne, Victoria 3122, Australia}
\author{Jesper.S. Hansen}
\email{jschmidt@ruc.dk}
\affiliation{``Glass and Time'', IMFUFA,
Department of Science and Environment, Roskilde University, Roskilde 4000, Denmark}

\date{\today}

\begin{abstract}
In this work, we investigate the effect of the hydrodynamic wall-fluid friction in electro-osmotic flows. First, we present the solution to the electro-hydrodynamic equation for the electro-osmotic velocity profile, which is derived for an ionic system composed of cations immersed in uncharged solvent particles. The system (solution and walls) is kept electrically neutral using negatively charged walls and will here be referred to as a ``counterion-only'' system. The theory predicts the existence of a counterion concentration that results in a maximum electro-osmotic flow rate, but only if the wall-fluid friction, or equivalently the slip length, is correlated with the system electrostatic screening length. Through equilibrium molecular dynamics simulations we precisely determine the hydrodynamic slip from the wall-fluid friction, and this is then used as input to the theoretical predictions. Comparison between the theory and independent non-equilibrium molecular dynamics simulation data confirms the existence of the maximum. Also, we find that standard hydrodynamic theory quantitatively agrees with simulation results for charged nanoscale systems for sufficiently small charge densities and ion charges, if the correct slip boundaries are applied. 
\end{abstract}

\pacs{}

\maketitle 


\section{Introduction}
Efficient fluid transport forms a critical component in the design of any micro or nanoscale fluidic system and has direct implications in a wide array of fields such as drug delivery, water desalination, pumping, and energy storage
\cite{lieber2003nanoscale, service2006desalination, ghosh2003carbon, siwy2002fabrication, zhou2013vibrating}.
In contrast to macroscale systems, conventional pressure-driven flow at the nanoscale is often impossible to achieve for practical applications due to the extremely high pressure gradient required to generate a net fluid flow inside the nanochannel \cite{hansen_2022}.
Hence, viable alternatives to pressure-driven flow have to be considered to transport fluid for nanoscale systems. Among various methods, electro-osmotic flow (EOF) has shown great promise in being an energy-efficient way to transport the fluid through nanoconfined systems \cite{alizadeh2021electroosmotic}, and thus understanding electro-osmotic flow behavior at the nanoscale can help in the efficient design of miniature devices.

Modeling electro-osmotic flows through nanochannels is often mathematically challenging because the classical assumptions usually employed in continuum theories may not capture the fluid behavior close to the walls. For example, the classical Poisson-Boltzmann (PB) equation cannot predict the ionic layering observed in the wall-fluid interfacial region, which can become highly influential when the channel size measures only a few nanometers \cite{qiao2003ion}. Another approximation that also requires serious consideration when studying nanoscale EOFs is the reliability of the no-slip boundary condition at the walls. Experimental and simulation studies have shown that the no-slip boundary condition usually employed in macroscale flows becomes insufficient to describe the hydrodynamics in nanochannels \cite{hansen_2022,hansen2011nanoflow,hansen2015continuum,karniadakis2006microflows}.
Hence at the nanoscale, it becomes essential to incorporate the fluid-solid interfacial slip in the theoretical prediction of the electro-osmotic velocity profile.

Previous studies have addressed the inadequacies in the prediction of EOF in nanochannels by introducing modifications to the classical PB equation and coupling with the hydrodynamics \cite{qiao2003ion, joly2004hydrodynamics, joly2006liquid}.
Even though appreciable progress has been made in modeling the electrostatics part in nanoconfined charged systems, precise and independent calculation of the hydrodynamic slip at the fluid-solid interface of a charged system still remains.
Most studies on nanoscale EOF have quantified the hydrodynamic slip using the parameter known as slip length \cite{joly2004hydrodynamics, huang2008aqueous, smiatek2008tunable, smiatek2009mesoscopic, bonthuis2012unraveling, uematsu2018analytical, silkina2019electro}.
This is often calculated from the velocity profile generated from nonequilibrium molecular dynamics (NEMD) simulations.
Even though NEMD simulations can provide an estimate for the effective slip for simple systems, this method of slip estimation does have some serious limitations.
First, the slip length calculated from NEMD simulations should be independent of the applied external driving field. This requires performing several simulations at different field strengths to determine the field-independent slip length, which can be a computationally intensive procedure \cite{thompson1997general}.
Second, for high slip systems, it has been shown that small uncertainties in the velocity profile can lead to large changes in the slip length \cite{kumar2012slip}, making the calculation of a precise value for hydrodynamic slip in this way highly questionable. 

Here, the EOF is investigated for an ionic solution with only a single type of dissolved ion confined between walls that have opposite charge to the ion.
Such a system is often referred to as a ``counterion-only'' system due to the absence of co-ions \cite{israelachvili2011intermolecular}. The system corresponds to a salt-free solution with charged walls and can be used to model realistic situations such as water flow through lipid bilayers, clay sheets, or along surfaces with ionizable sites, and also in the case of lamellar liquid crystals \cite{israelachvili2011intermolecular, engstrom1978ion}. The hydrodynamic slip is quantified using the interfacial friction coefficient (also known as the Navier friction coefficient), which, to overcome the problems with NEMD estimates mentioned above, is computed from the equilibrium molecular dynamics simulations using the methodology proposed in the work of Hansen et al.\cite{hansen2011prediction} and  Varghese et al. \cite{varghese2021improved}.
To understand the effect of coupling between electrostatics and hydrodynamic slip, we derive an approximate analytical solution for the volumetric flow rate of a counterion-only system.
We find that there exists an optimal counterion concentration that can result in a maximum electro-osmotic flow rate in the case of non-wetting nanochannels. This is confirmed by the NEMD simulations. 

\section{Theory}
\begin{figure}
\includegraphics[width=1.0\textwidth]{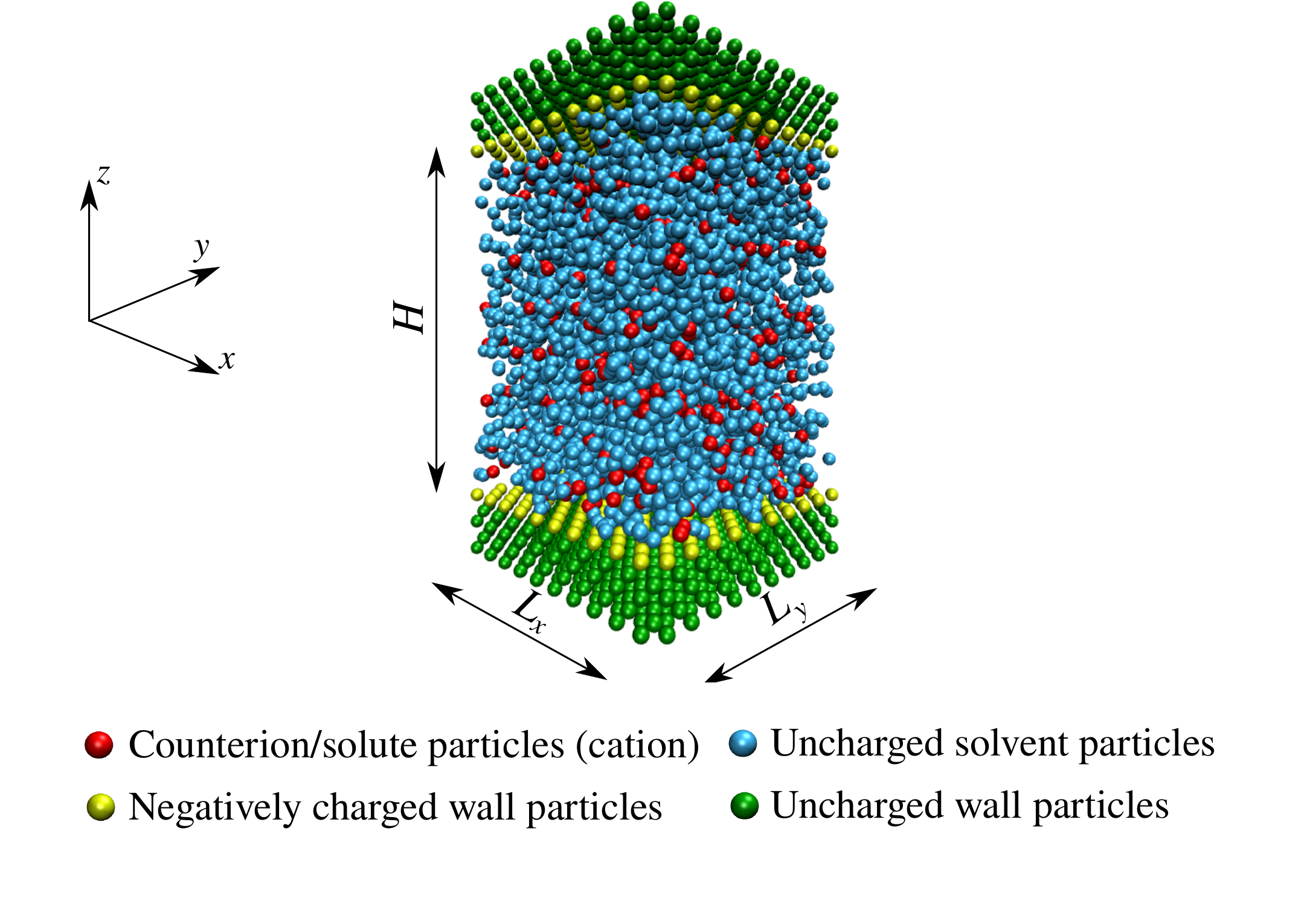}
\caption{Molecular representation of the charged system used in this study.
The walls are placed in the $x-y$ plane with a wall area equal to $L_{x} \times L_{y}$. The channel height, $H$, is defined as the available channel height for the mobile ions; this is estimated from the density and velocity flow profiles.
}
\label{fig:sys}
\end{figure}

We start our theoretical treatment from the Poisson-Boltzmann (PB) equation for a counterion-only system. For a nano-slit pore with confinement along the $z$-axis, as shown in Figure \ref{fig:sys}, the one-dimensional form of the PB equation is given as \cite{bruus2008theoretical}
\begin{eqnarray}
\frac{d^2\psi}{dz^2} = -\frac{q\rho_{0}}{\varepsilon}
\textrm{e}^{-q\psi/k_{B}T}, \label{eq:pb_1d}
\end{eqnarray}
where $z$ is the spatial coordinate, and we choose the coordinate system such that  $z=0$ is the mid-point of the channel. $\psi$ is the electric potential,  $\rho_{0}$ is the reference ion number density, where the electric potential becomes zero, $q$ is the charge of the ions, $k_{B}$ is the Boltzmann constant, $T$ is the system temperature, and $\varepsilon = \varepsilon_{0}\varepsilon_{r}$, where $\varepsilon_{0}$ and $\varepsilon_{r}$ are the vacuum permittivity and relative permittivity, respectively. We will assume that the charge dependency of $\varepsilon$ can be ignored. 
Note, the channel height, $H$, is the distance between the center of mass of the inner-most wall layers minus the hydrodynamic stationary (stagnant) Stern layer if this is present.

An analytical solution exists for Eq. (\ref{eq:pb_1d}) and is given by \cite{israelachvili2011intermolecular}
\begin{eqnarray}
\psi(z) = \frac{k_{B}T}{q}\textrm{ln}\big(\textrm{cos}^{2}(z/\lambda)\big),
\label{eq:electric_pot_israelachivili}
\end{eqnarray}
where $\lambda$ is the screening length parameter 
\begin{eqnarray}
\lambda = \sqrt{\frac{2\varepsilon k_{B}T}{q^{2} \rho_{0}}}. \label{eq:lambda}
\end{eqnarray}
From the solution, Eq. (\ref{eq:electric_pot_israelachivili}), we can see that the reference electric potential is $\psi = 0$ at $z=0$, such that $\rho_{0}$ corresponds to the counterion concentration at the channel center, that is, $\rho_{0} = \rho(0)$. Importantly, enforcing continuity on the solution Eq. (\ref{eq:electric_pot_israelachivili}) there exists a lower bound $0<\cos^2(z/\lambda)$ implying that if we define the half-height $h=H/2$, we have the constraint $\lambda > 2 h/\pi$ since $-\pi/2 < z/\lambda < \pi/2$ and $-h \leq z \leq h$. This means that Eq. (\ref{eq:electric_pot_israelachivili}) is only valid for sufficiently large screening lengths; this is true in the Debye-H\"{u}ckel limit $k_BT \gg q^2\rho_0$. 

Now, using Eq. (\ref{eq:electric_pot_israelachivili}) and the Boltzmann distribution for counterions, $\rho(z) = \rho_{0} \textrm{e}^{-q\psi(z)/k_{B}T}$,
we obtain the counterion concentration profile in terms of the screening length and reference concentration
\begin{eqnarray}
\rho(z) = \frac{\rho_0}{\textrm{cos}^{2}(z/\lambda)} \label{eq:counterion_density_profile}.
\end{eqnarray}
Applying the charge neutrality condition we have the relation
\begin{eqnarray}
-2\Sigma_{\textrm{wall}} &=& \int_{-h}^{h} q\rho(z)dz, \label{eq:electro_neutrality}
\end{eqnarray}
where $\Sigma_{\textrm{wall}}$ is the surface charge density of one wall. Substituting Eq. (\ref{eq:counterion_density_profile}) in Eq. (\ref{eq:electro_neutrality})
we can get a relation between the screening length and surface charge density of the wall as 
\begin{eqnarray}
\frac{2k_{B}T}{q\lambda}\textrm{tan}(h/\lambda) = -\frac{\Sigma_{\textrm{wall}}}{\varepsilon}.\label{eq:find_lambda_israel}
\end{eqnarray}
Eq. (\ref{eq:find_lambda_israel}) can be further simplified using Eq. (\ref{eq:lambda}) 
to the form 
\begin{eqnarray}
\lambda \textrm{tan}(h/\lambda) = -\frac{\Sigma_{\textrm{wall}}}{q\rho_0}.
\label{eq:find_lambda_modified}
\end{eqnarray}
Since we do not explicitly compute the 
permittivity we apply Eq. (\ref{eq:find_lambda_modified}) to calculate 
$\lambda$ with $\rho_0$ as an input parameter from molecular dynamics simulations.

Once the ion concentration distribution is known we can write the Navier-Stokes equation for the EOF. For low Reynolds number and in the steady-state this is \cite{bruus2008theoretical, hansen_2022}
\begin{eqnarray}
\eta_{0} \frac{d^{2}u_{x}}{dz^{2}} =  -q \rho E_{x}, \label{eq:eof_eq}
\end{eqnarray}
where $u_{x}$ is the $x$-component of the velocity, $\eta_{0}$ is the shear viscosity and $E_{x}$ is the external electric field applied in the $x$-direction. We assume that $\eta_0$ is independent of the charge concentration. Substitution of Eq. (\ref{eq:counterion_density_profile}) in Eq. (\ref{eq:eof_eq}) leads to
\begin{eqnarray}
\frac{d^{2}u_{x}}{dz^{2}} = -\frac{q\rho_0\ E_x}{\eta_{0}\ \textrm{cos}^{2}(z/\lambda)}, \label{eq:eof_eq2}
\end{eqnarray}
and integrating yields
\begin{eqnarray}
u_{x}(z) = \frac{q\rho_0\lambda^{2} E_x}{\eta_0} \ln \left(\textrm{cos}(z/\lambda)\right)\ E_{x}+ D_{1}z + D_{2} \, , 
\label{eq:eov_gen_soln}
\end{eqnarray}
where $D_1$ and $D_2$ are constants of integration. We use the system symmetry and slip boundary conditions, i.e.,
\begin{eqnarray}
u_{x}(-h) = u_{x}(h) = u_{s}, \label{eq:eov_sym_bc}
\end{eqnarray}
where $u_{s}$ is the slip velocity, and we arrive at the solution for the velocity profile
\begin{eqnarray}
u_{x}(z) = \frac{q\rho_0\lambda^{2}E_{x}}{\eta_{0}}
\ln \left[\frac{\textrm{cos}(z/\lambda)}{\textrm{cos}(h/\lambda)}\right]  + u_{s}. \label{eq:eov_final_soln}
\end{eqnarray}
We wish to express the slip velocity in terms of the slip-length, $L_s$. From the continuity of the shear pressure (or stress) at the interface between the mobile fluid and stagnant surface we have \cite{hansen_2022} 
\begin{eqnarray}
u_{s} = \frac{\eta_{0}}{\xi_{0}}\frac{du_{x}(z)}{dz}\bigg|_{z=\pm h}
= \frac{q\rho_0\lambda E_{x}\ \textrm{tan}(h/\lambda)}{\xi_{0}} \, , \label{eq:slip_vel}
\end{eqnarray}
where $\xi_{0}$ is the interfacial friction coefficient. Note that Eq. (\ref{eq:slip_vel}) is the Navier-slip boundary condition, which is simply a Neumann boundary. By combining Eqs. (\ref{eq:eov_final_soln}) and (\ref{eq:slip_vel}), we can write the EOF velocity in the final form
\begin{eqnarray}
u_{x}(z) = \frac{q\rho_0\lambda^{2} E_x}{\eta_0}\left(
\ln \left[\frac{\textrm{cos}(z/\lambda)}{\textrm{cos}(h/\lambda)}\right]  + \frac{L_s \tan(h/\lambda)}{\lambda} 
\right) \, ,
\nonumber \\ \label{eq:eov_rearrange}
\end{eqnarray}
where $L_{s}=\eta_{0}/\xi_{0}$ is the slip length. Eq. (\ref{eq:eov_rearrange}) represents an electro-hydrodynamic model for the velocity profile, where the second term on the right-hand describes the effect of the hydrodynamic slip.

The flow can be quantified by the volumetric flow rate, $Q$, through a cross-section of the pore, $[-h,h] \times [-w, w]$, by
\begin{equation}
\label{eq:flowrate}
    Q = \int_{-w}^w dy \int_{-h}^h u_x(z) \, dz = 2w \int_{-h}^h u_x(z) \, dz \, .
\end{equation}
To our knowledge, the integral does not have a known solution of elementary functions, and therefore we Taylor expand around the channel center, $z=0$, approximating $\ln|\cos(z/\lambda)| \approx  -z^2/(2\lambda^2)$. Equation (\ref{eq:flowrate}) is then 
\begin{equation}
    \label{eq:Qapprox}
    Q \approx -\frac{2q\rho_0\lambda^{2} E_x}{\eta_0} \left[
    \frac{h^3}{3\lambda^2} + 2h\ln\left(\cos(h/\lambda)\right) - 
    \frac{2L_sh}{\lambda} \tan(h/\lambda)
    \right]  \, .
\end{equation}
First, consider the flow rate in the special case of no-slip, $L_s=0$. We then investigate the flow rate dependency of the screening length, $Q=Q(\lambda)$. Since we have the relation
\begin{eqnarray}
-\frac{d}{d\lambda} \left(
    \frac{h^3}{3\lambda^2} + 2h\ln\left(\cos(h/\lambda)\right) \right) =
       \frac{2}{3}\left(\frac{h}{\lambda}\right)^3 - 2 \left(\frac{h}{\lambda}\right)^2 \tan(h/\lambda) < 0 
\end{eqnarray}
for $2h/\pi < \lambda < \infty$, we see that the flow rate decreases monotonically with screening length, thus, the maximum flow rate is obtained in the limit 
$\lambda \rightarrow 2h/\pi$ for zero slip. 

\begin{figure}
    \includegraphics[width=0.45\textwidth]{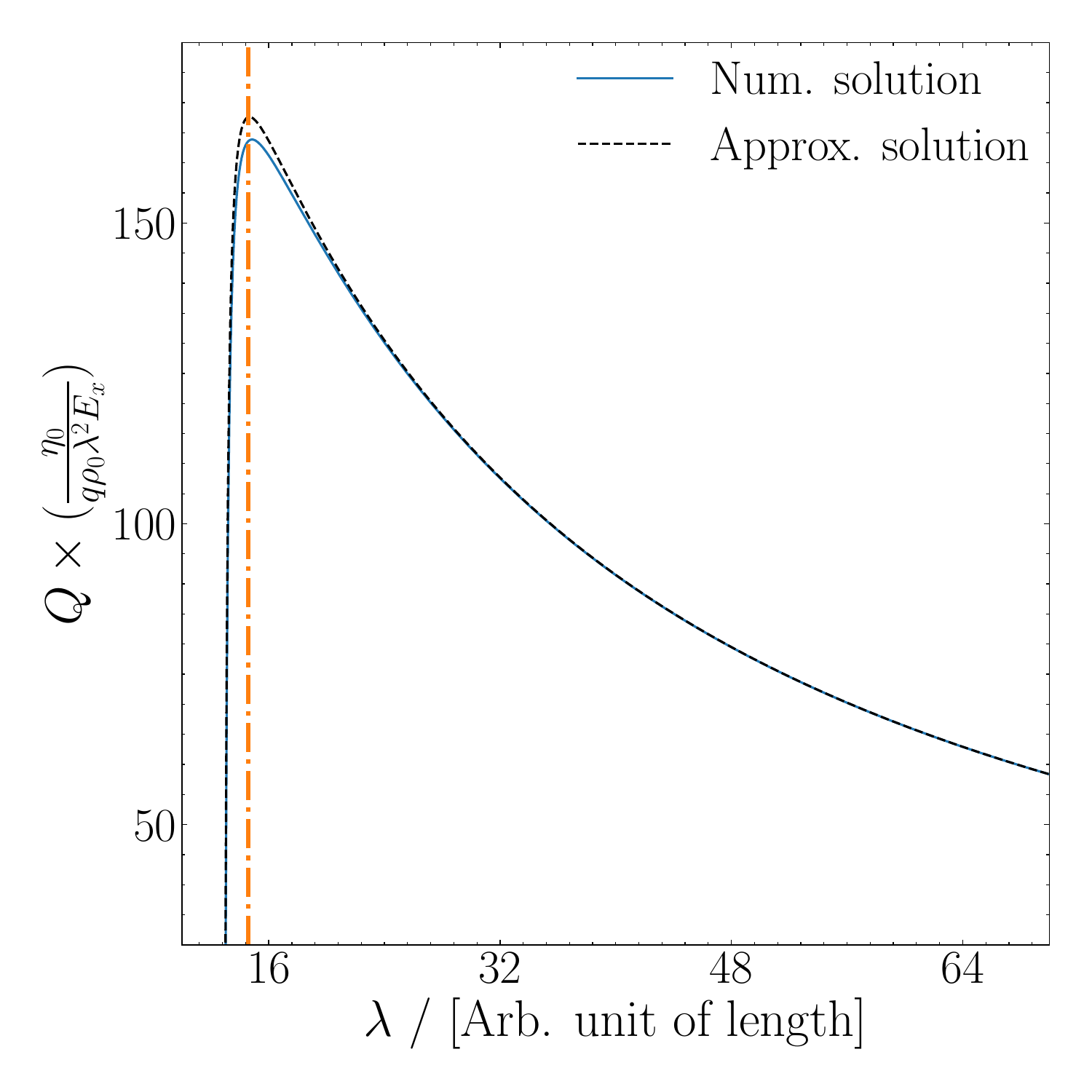}
    \caption{
        Volumetric flow rate, $Q$ as a function of screening length, $\lambda$, 
        using the correlation function $L_s = \alpha \lambda - \beta$. In arbitrary units we have applied variable values $h=20$, $\alpha = 6$, $\beta = 13\alpha + 1=79$, and $13 \leq \lambda \leq 100$. Vertical line indicates the value for $\lambda_\text{max}$ obtained from Eq. (\ref{eq:lambda0}).
    }
    \label{fig:Q}
\end{figure}

Now, consider the case where the slip length is non-zero. One must expect that the slip length depends on the detailed ion layering in the wall-fluid interface region. This in turn depends on the ion concentration or equivalently the screening length, and therefore we have a correlation between the slip length and the screening length, $L_s = L_s(\lambda)$. By differentiation of Eq. (\ref{eq:Qapprox}) we then get the implicit equation for the possible maximum flow rate at $\lambda=\lambda_\text{max}$
\begin{eqnarray}
    \label{eq:lambda0}
    \frac{2}{3}\left(\frac{h}{\lambda_\text{max}}\right)^3 - \frac{2h}{\lambda_\text{max}}\left(
        \frac{L_s+h}{\lambda_\text{max}} - L_s'(\lambda_\text{max})  
    \right)\tan(h/\lambda_\text{max})  - \nonumber \\ \frac{2h^2L_s(\lambda_\text{max})}{\lambda_\text{max}^3}\sec^2(h/\lambda_\text{max}) = 0 \nonumber  \\
\end{eqnarray}
The correlation function is unknown a priori. As a simple example we here choose the form $L_s = \alpha \lambda - \beta$, where $\alpha > 0$, $\lambda > 2h/\pi$ (by the constraint discussed above), and $\beta$ is chosen such that $L_s \geq 0$, that is, $\beta \leq \alpha \lambda$. Figure \ref{fig:Q} shows the volumetric flow rate for this choice of correlation function: full line is the result obtained from direct numerical integration of Eq. (\ref{eq:eov_rearrange}), and the dashed line from the approximation Eq. (\ref{eq:Qapprox}). We see that $Q$ indeed features a maximum at $\lambda_\text{max} \approx 14.6$ (in some arbitrary unit) as predicted by Eq. (\ref{eq:lambda0}) (vertical line); here a standard minimization algorithm is used to find the value for $\lambda_\text{max}$ \cite{octave:manual}.

\section{Molecular dynamics simulations}
Molecular dynamics simulations were performed using the large atomic/molecular massively parallel simulator (LAMMPS) package \cite{plimpton1995fast}. All particles (both fluid and wall) in the system were modeled as simple spherical particles that interact pair-wise via the Lennard-Jones (LJ) potential 
\begin{eqnarray}
\phi(r) =
\begin{cases}
4 \epsilon\Big[\Big(\frac{\sigma}{r}\Big)^{12}-C\Big(\frac{\sigma}{r}\Big)^{6}\Big]
& \textrm{if} \ r \leq r_c, \\
0 & \textrm{if} \ r > r_c,
\end{cases} \label{eq:lj}
\end{eqnarray}
where $\sigma$ and $\epsilon$ are the LJ parameters defining the simulation characteristic length and energy scales, respectively. $r$ is the interatomic distance between the atoms, and $r_c$ is the interaction cut-off, which is kept at $2.5\sigma$. $C$ is the wetting coefficient that controls the different wall-fluid wetting properties; we used a fluid-fluid wetting coefficient of $C = 1.2$, and for the wall-fluid, we used a non-wetting value $C = 0.5$. Notice that the wetting can also be controlled by the LJ parameters $\sigma$ and $\epsilon$ \cite{alosious2019prediction}. All particles have the same mass, $m$, and the mechanical properties can be expressed in terms of $\sigma$, $\epsilon$, and $m$. As it is common practice we will from hereon set the parameters to unity and omit these units as well as the units relating to the electrostatics. 

The short-range Coulombic interactions were calculated using an interaction cut-off of $r_c = 2.5$, and for the long-ranged interactions the Ewald algorithm is used with the particle-particle-particle-mesh solver of LAMMPS \cite{hockney1981computer,plimpton1995fast,yeh1999ewald} with a relative root mean square error in the per-atom force calculations below $1 \times 10^{-4}$. To accommodate the non-periodicity in the $z$-direction, we used the corrected Ewald algorithm EW3DC \cite{yeh1999ewald}, where the ratio of extended volume to actual channel size is set as 3.0.

We studied systems with two different ion charges, $q=+0.2$ and $q=+1.6$, and with varying ion concentrations. All systems have the same overall fluid density of $0.9$. Tables \ref{tab:sys_info_q02} and \ref{tab:sys_info_q16} list the different 
system properties for $q=+0.2$ and $q=+1.6$, respectively; all quantities are given in LJ units as mentioned above, however, it is important to note that for a typical value for the length scale, say $\sigma = 0.5$ nm, the molarity of the systems ranges from 0.08 M to 7 M, that is, the concentration range falls within realistic values. 

The walls were placed in the $x-y$ plane with periodicity in both directions and symmetry about the channel center, $z=0$. Each wall consists of five layers of atoms arranged in a face-centered cubic lattice with a density of 1.0 and an interlayer distance of $0.8$. The wall particles were maintained around their initial equilibrium positions $\textbf{r}_{\textrm{eq}}$, using a harmonic potential given by, $\phi_{s} = \frac{1}{2} k_{s} (\textbf{r}_{i}-\textbf{r}_{\textrm{eq}})^2$, where $k_s = 150$ is the spring constant and $\textbf{r}_{i}$ is the instantaneous position of the wall particle \cite{todd2017nonequilibrium}. The inner-most layer of each wall was negatively charged to maintain electroneutrality for the system, with the total charge on each wall being $-qN_{\textrm{ci}}/2$, where $N_{\textrm{ci}}$ is the total number of counterions in the system, see Table \ref{tab:sys_info_q02}. The wall area in the $x-y$ plane is, $L_{x} \times L_{y} = 16 \times 16 = 256$, where $L_{x}$ and $L_{y}$ are the simulation box dimensions along the $x$ and $y$ axes respectively. Figure \ref{fig:sys} shows a snapshot of the system.

Equilibrium molecular dynamics (EMD) simulations for each system were performed for $3 \times 10^{6}$ steps in total, where the initial $2 \times 10^{6}$ steps were used to equilibrate the system at  $T = 1.0$ by thermostating the walls using a
Nos\'e-Hoover \cite{nose1984unified,hoover1985canonical} thermostat.
The final $1 \times 10^{6}$ steps were used to collect data for post-processing.
The equations of motion for all particles were integrated using the
Velocity-Verlet\cite{swope1982computer} scheme with an
integration time step $\Delta t = 0.001$. 20 independent EMD simulations were performed for each system to obtain sufficient statistics.

We also performed nonequilibrium molecular dynamics (NEMD) simulations to investigate the EOF for each system. Each NEMD simulation was performed for a total of $4 \times 10^{6}$ steps, with the initial $2 \times 10^{6}$ steps used for equilibration of the system at $T = 1.0$, by thermostating the walls using the Nos\'e-Hoover \cite{nose1984unified,hoover1985canonical} thermostat.
After the initial equilibration, an external electric field, $E_{x} = $ 0.075 was applied for the final $2 \times 10^{6}$ steps to generate the flow, where the data was collected from the final $1 \times 10^{6}$ steps.
Every NEMD result was also averaged across 20 independent simulations.

\section{Results and Discussion}
\begin{figure}
\includegraphics[width=0.4\textwidth]{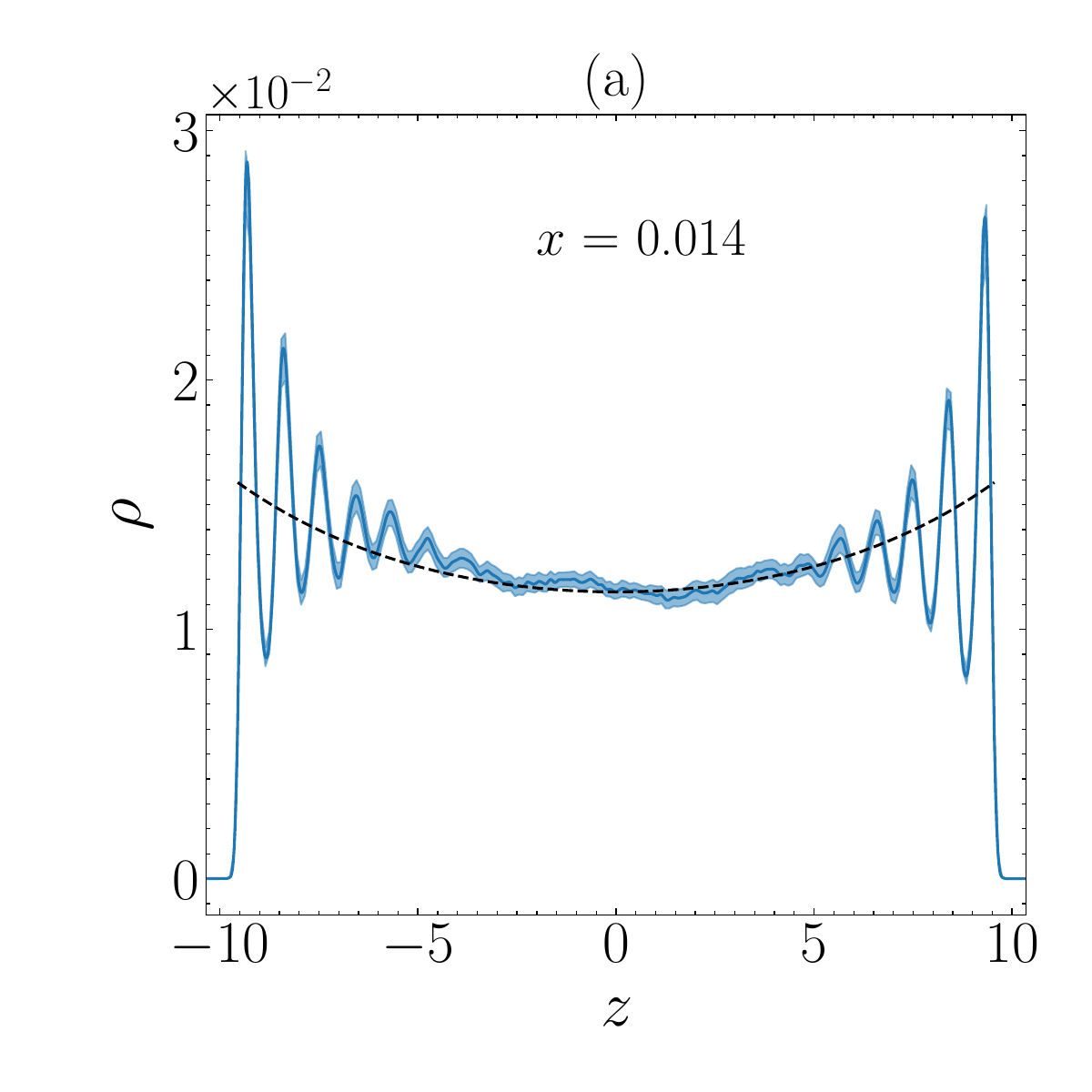}\\
\includegraphics[width=0.4\textwidth]{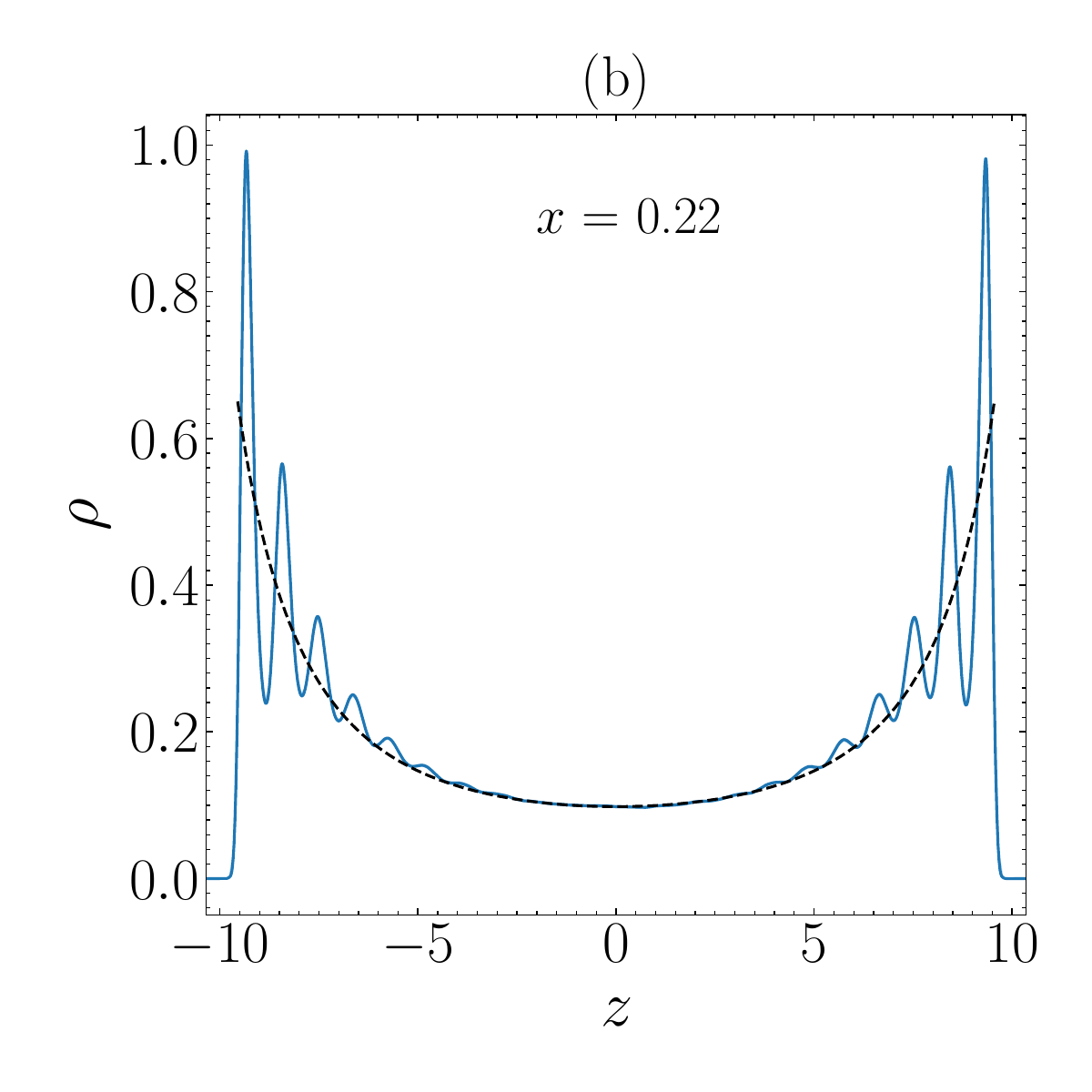}\\
\includegraphics[width=0.4\textwidth]{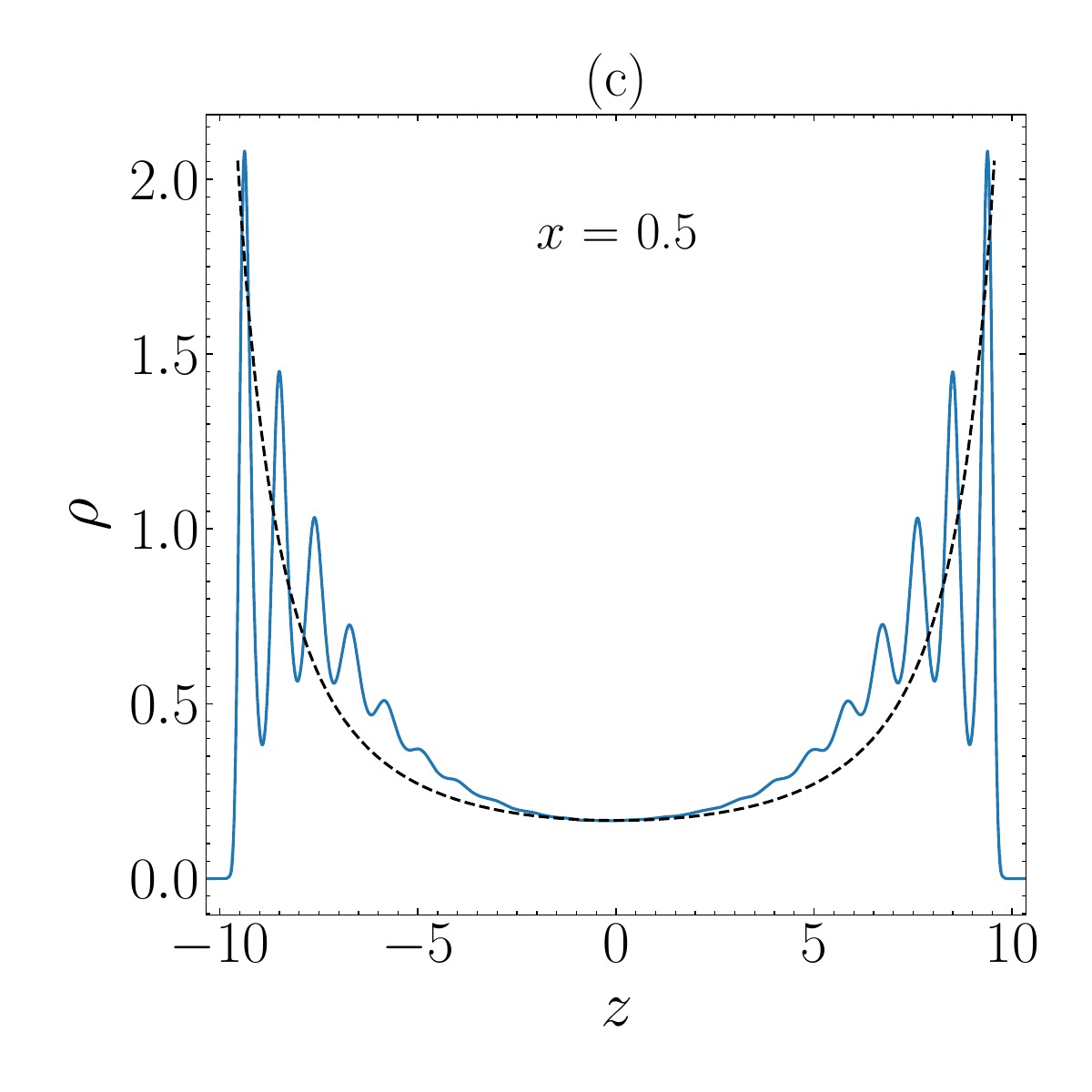}\\
\caption{Variation of the counterion density $\rho$, for different counterion mole fractions across the channel width. The solid lines represent the density profile obtained from EMD simulations, and the dashed lines represent the density profile predicted by Eq. (\ref{eq:counterion_density_profile}).
The shaded region in (a) corresponds to the statistical error associated with the data.
}
\label{fig:counterion_density_profile}
\end{figure}

First, we focus on the case with ion charge $q=+0.2$. Figure \ref{fig:counterion_density_profile} shows the EMD ion distribution (or profiles) predicted by Eq. (\ref{eq:counterion_density_profile}) (dashed line) and that obtained from the simulations (full line). In this comparison, the value for $\rho_0$ is determined directly from the simulations, and the screening length is calculated numerically from the relation in Eq. (\ref{eq:find_lambda_modified}), see values in Table \ref{tab:sys_info_q02}. Since the statistical uncertainty associated with $\rho_{0}$ is negligible, we do not consider the propagation of this statistical uncertainty in our computation of $\lambda$.

\begin{table}
\caption{Summary of the calculated system properties for ion charge $q=+0.2$. $\rho_0$ the ion density at $z=0$, $\lambda$ is the screening length, $\xi_0$ the friction coefficient, and  $Q$ the measured flow rate from the NEMD simulations. 
Except for $Q$, all properties are calculated from EMD simulations. $h=H/2$ is estimated to be 9.7 for all systems. Notice that for $x=0.61$ the friction coefficient cannot be calculated and we therefore do not estimate the other system properties.}
\begin{ruledtabular}
\begin{tabular}{cccccc}
\noalign{\vskip 2mm} 
$x$ & $\rho_0$ & $\lambda$ & $\xi_0$ & $\Sigma_\text{wall}$ & $Q$ (NEMD)\\
\noalign{\vskip 2mm} 
\hline 
\noalign{\vskip 2mm} 
0.007   & 0.006 & 28.92 &  1.27 $\pm$ 0.03 & -0.0125 &  0.4 $\pm$ 0.3 \\
0.014   & 0.011 & 17.26 &  1.25 $\pm$ 0.03 & -0.025 & 0.7 $\pm$ 0.2 \\
0.049   & 0.036 & 12.72 &  1.30 $\pm$ 0.03 & -0.0875 & 3.1 $\pm$ 0.3 \\
0.11    & 0.064 & 9.55 &  1.35 $\pm$ 0.03 & -0.2 & 6.2 $\pm$ 0.2 \\
0.22    & 0.098 & 8.15 &  1.78 $\pm$ 0.05 & -0.4 & 10.3 $\pm$ 0.2 \\
0.33    & 0.122 & 7.14 &  2.45 $\pm$ 0.07 & -0.6 & 12.0 $\pm$ 0.2 \\
0.39    & 0.133 & 7.51 &  3.08 $\pm$ 0.07 & -0.7 & 12.2 $\pm$ 0.2 \\
0.50    & 0.166 & 7.45 & 5.43 $\pm$ 0.13  & -0.9 & 12.1 $\pm$ 0.2 \\
0.61    &   -   &  -   &     -             &  -    & 7.1  $\pm$ 0.3 \\
\noalign{\vskip 2mm} 
\end{tabular}
\end{ruledtabular}
\label{tab:sys_info_q02} 
\end{table}

Recall, that the channel width is here defined to be the part of the channel which is available to the mobile fluid particles. From the density profile, we estimate $H \approx 19.4$; this value is independent of whether the external force is applied or not. Since $h=H/2$ enters the expression for the EOF velocity profile, the predicted profile and flow rate will indeed depend on this estimate. The authors are not aware of an unambiguous definition \cite{hansen_2022}.    

Clearly, Eq. (\ref{eq:counterion_density_profile}) does not capture the layering of the counterions close to the walls, but for sufficiently small ion concentrations it shows satisfactory prediction for the density in the bulk region of the channel. The external force density that drives the flow, $q\rho E_x$, will vary strongly in the wall-fluid region and is therefore not correctly modeled by Eq. (\ref{eq:counterion_density_profile}). Recently, it has been shown that incorporating the coupling between the solvent density and ion density into the PB theory, one can successfully model the layering, at least for low counterion concentrations \cite{joly2004hydrodynamics, joly2006liquid}. It is not the purpose here to pursue this coupling effect, but we will comment on the point below. 

To quantify the hydrodynamic slip, we compute the interfacial friction coefficient, $\xi_0$, using the equilibrium method proposed by Varghese et al. \cite{varghese2021improved}, which is a statistical improvement of the method devised by Hansen et al. \cite{hansen2011prediction}. In Supplemental Information it is shown how it is applied in this particular case of a charged system. Table \ref{tab:sys_info_q02} lists the friction coefficients for charge $q=+0.2$; we observe that increasing the ion concentration increases hydrodynamic friction between the wall and fluid particles, concurring with the results reported in previous studies\cite{joly2006liquid, xie2020liquid}. We also find that the method becomes unreliable above mole fractions $x > 0.5$; the exact reason for this limitation is not clear, however, we expect it to be due to the emergence of crystallization in the interfacial region, leading to unrealistic values for interfacial friction.

From $\xi_0$ the slip length, $L_s$, can be found using the definition $L_s = \eta_0/\xi_0$. To this end, we have performed a series of NEMD simulations in order to investigate the viscosity dependence of the ion concentration (see Supplementary Information),  and we conclude that the viscosity does not vary significantly in the concentration range studied here. This also validates the assumption in the theory section. Thus, the viscosity equals, within the standard deviation, the (non-charged solvent only) case viscosity $\eta_0 = 4.1 \pm 0.1$. 

\begin{figure}
\includegraphics[width=0.4\textwidth]{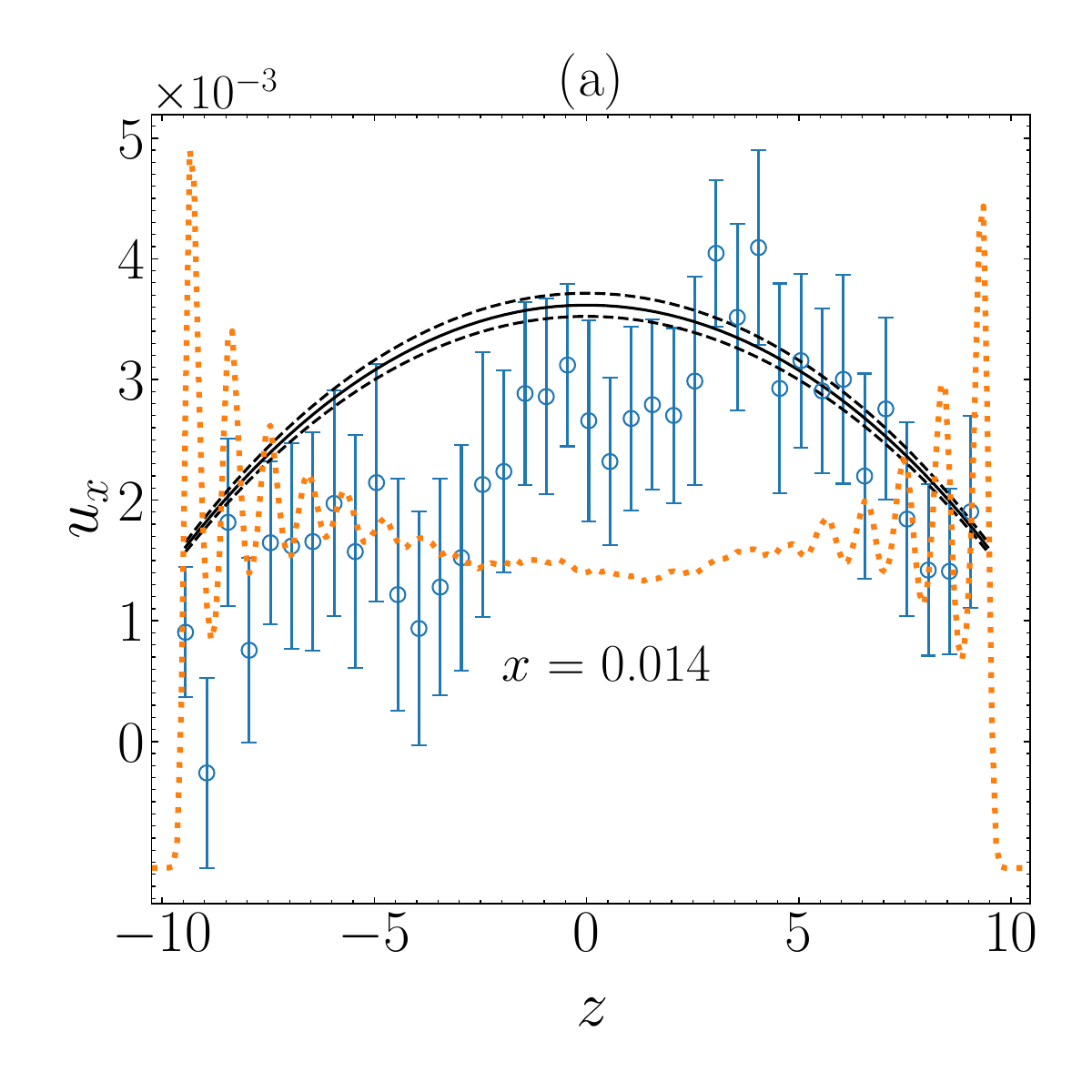}\\
\includegraphics[width=0.4\textwidth]{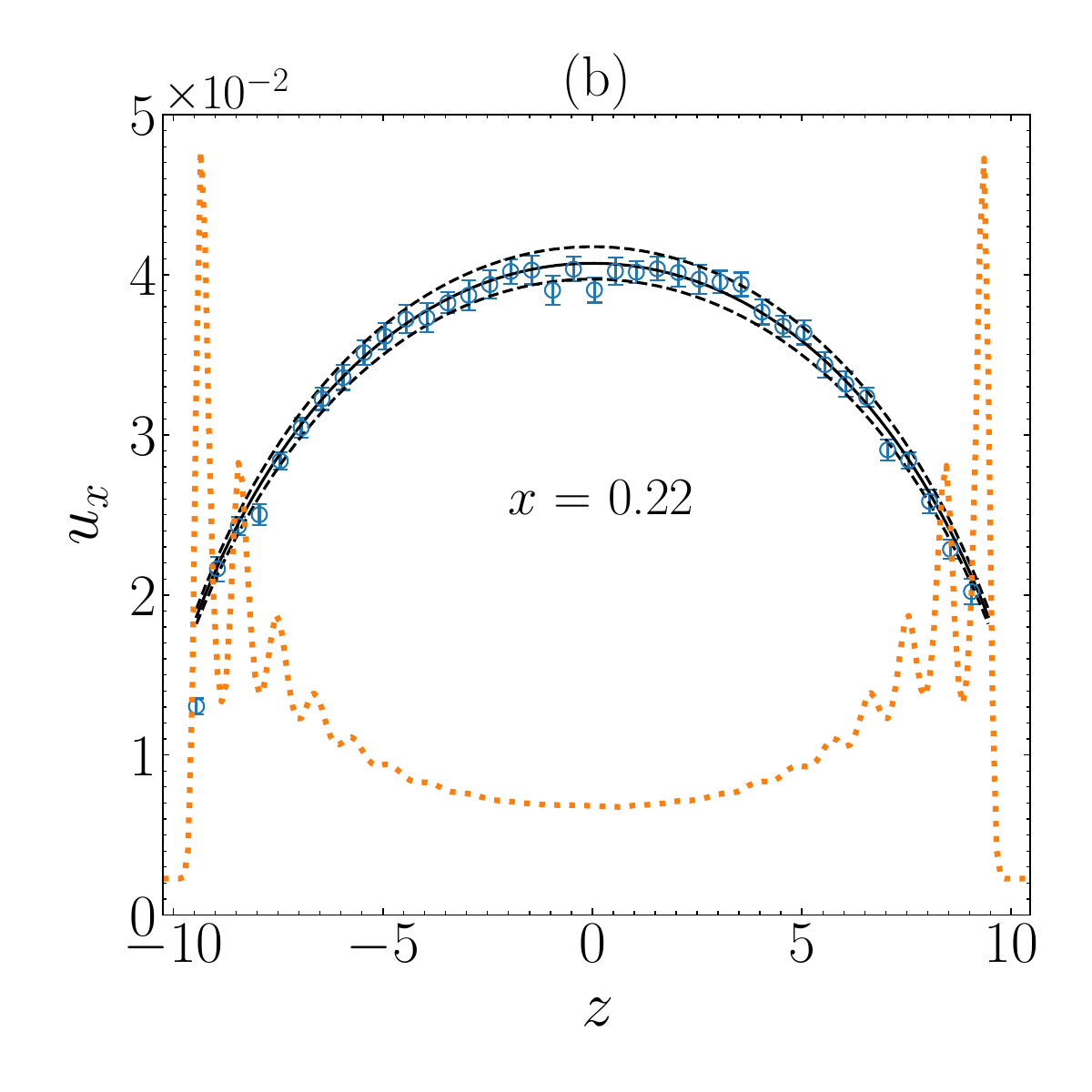}\\
\includegraphics[width=0.4\textwidth]{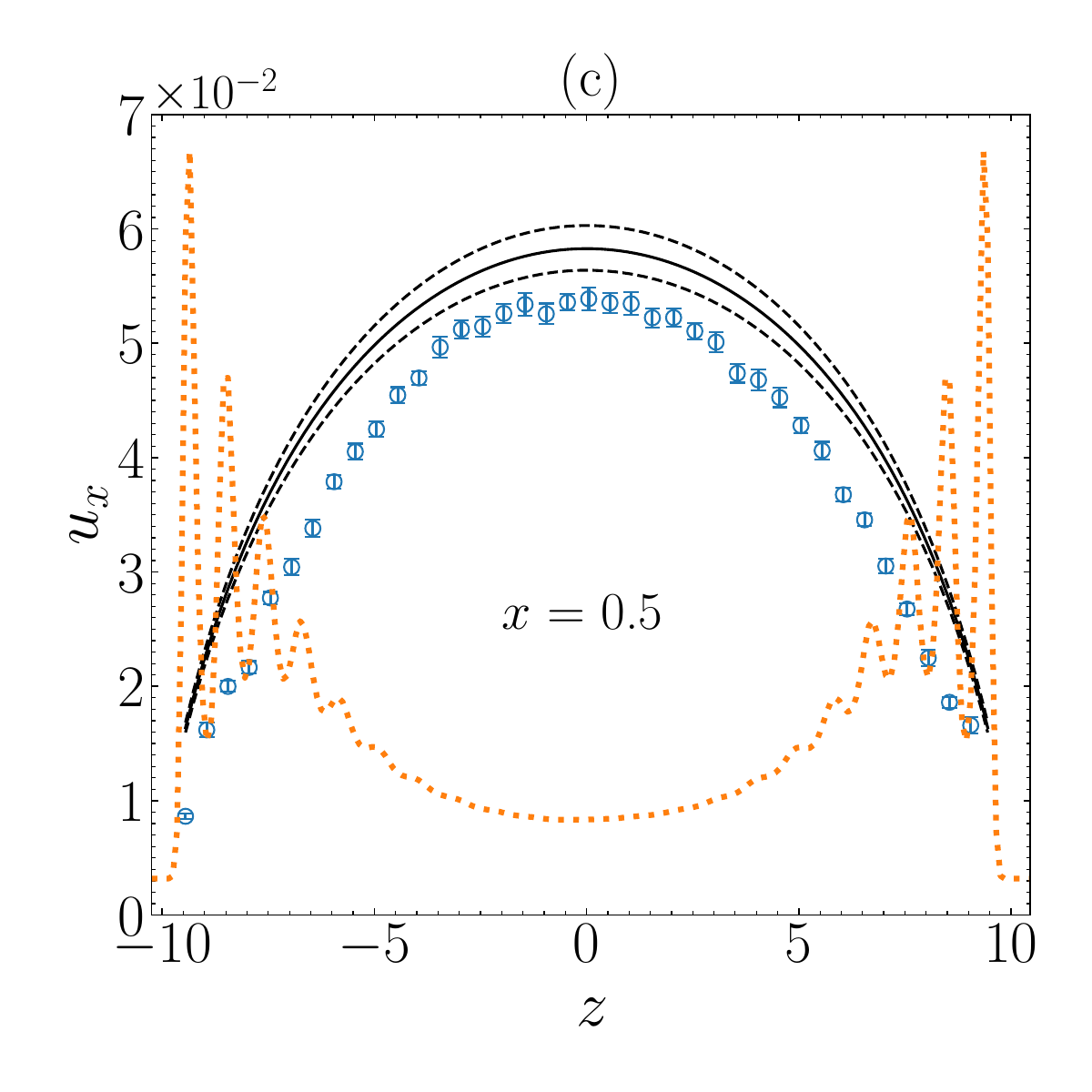}\\
\caption{Electro-osmotic velocity profile, $u_x$, for different counterion concentrations. Circles represent the electro-osmotic velocity profile obtained from NEMD simulations. Solid lines represent the electro-osmotic velocity profile predicted using Eq. (\ref{eq:eov_rearrange}). Dashed lines are the error range for the predicted profiles due to the standard errors in the values for $\eta_{0}$ and $\xi_{0}$. 
The superimposed dotted lines are scaled and shifted counterion density profiles. 
}
\label{fig:velprof}
\end{figure}

Next, we compare the velocity profiles obtained in the simulations with the predicted profiles, Figure \ref{fig:velprof}, using the parameter values in Table \ref{tab:sys_info_q02}, that is, in the predictions we do not perform any fitting to the NEMD profiles. Note, we have excluded predictions from the highest concentration, $x=0.61$, since the friction coefficient was not determined for this system.

For $x=0.014$ the noise-to-signal ratio is very large in the simulation data and the profile curvature is not clearly visible. Due to these large statistical uncertainties, the direct comparison is inconclusive for low concentrations. For intermediate concentrations, see Figure \ref{fig:velprof} (b), the standard theory agrees very well with the simulation data when the slip boundary is applied. We highlight again that no direct fitting is carried out in the comparison. There is no observable effect of the ion layering on the flow profile in the wall-fluid region. This phenomenon is equivalent to what is seen in the standard Poiseuille flow, where the flow profile is unaffected by the density layering for sufficiently large channels due to the non-local fluid response \cite{todd:pre:2008, hansen_2022}. Thus, even if the external driving force is given by the charge density the small-scale variation has no effect on the flow profile. For large ion concentrations, Figure \ref{fig:velprof} (c), one clearly sees that the ion layering affects the flow in the wall-fluid region. The fluid in the region is not stagnant, hence, it does not form a Stern layer, however, the strain rate is reduced compared to the predictions indicating a reduced shear stress in the wall-fluid region.   

\begin{figure}
\centering
\includegraphics[width=0.45\textwidth]{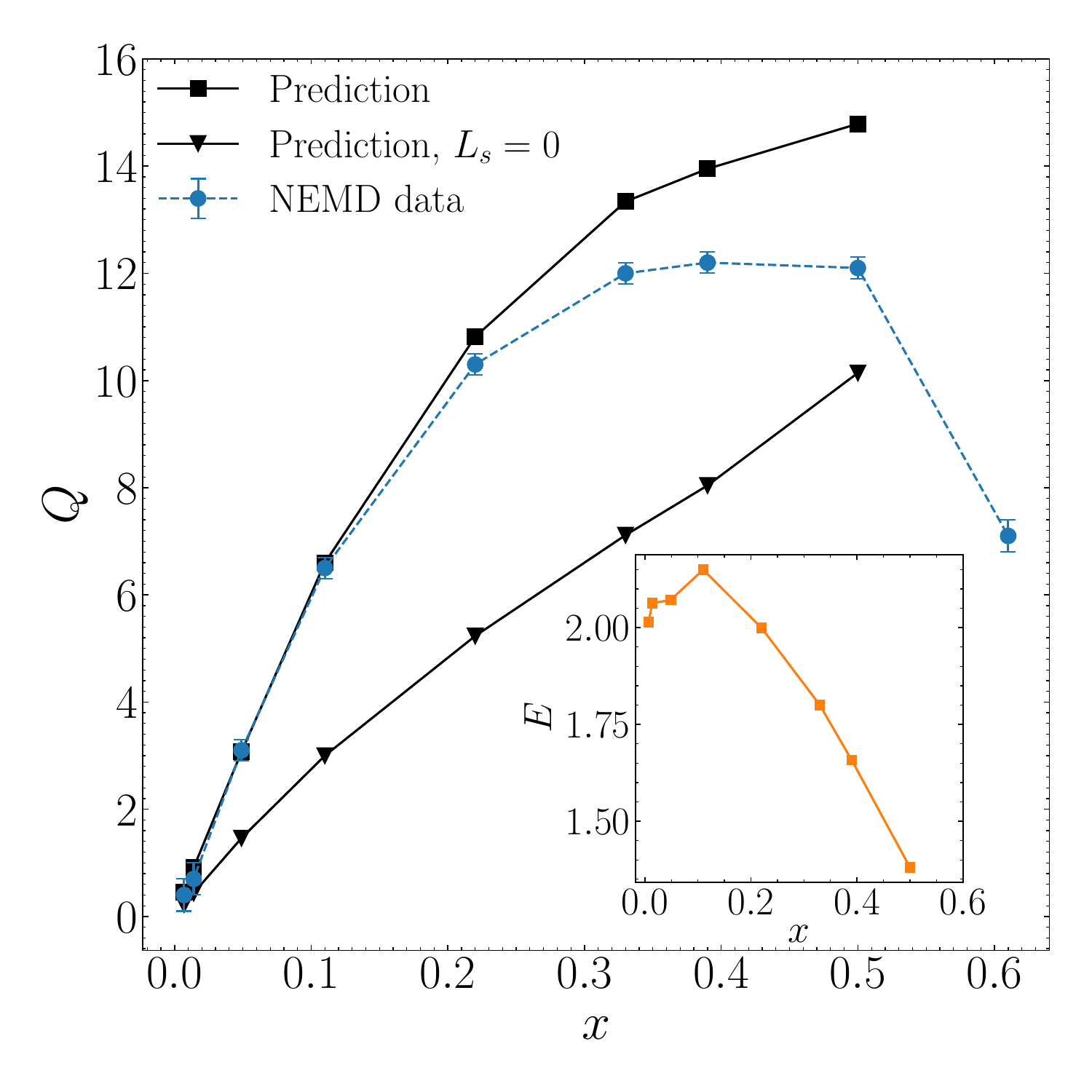} 
\caption{
Volumetric flow rate versus counterion concentration for $q = +0.2$. Circles connected with the dashed line are the results from the NEMD simulations, squares connected with the solid line are the theoretical predictions with slip and triangles connected with the 
solid-line are the theoretical predictions with no-slip. 
Inset: The variation of the enhancement coefficient, $E$, with counterion concentration.
}
\label{fig:flowratecmp}
\end{figure}

Figure \ref{fig:flowratecmp} shows the volumetric flow rate obtained from the simulations and the predicted flow rate, Eq. (\ref{eq:Qapprox}), using both zero and non-zero slip. The simulation flow rates are calculated directly from Eq. (\ref{eq:flowrate}) using $w=8$ and the velocity profiles from simulations. While Eq. (\ref{eq:Qapprox}) 
overestimates the flow rate for larger counterion concentrations, the prediction is in excellent 
agreement with simulation data at lower concentrations. 

We note that for $q=+0.2$ the simulation data for the flow rate features a plateau and then drops at $x=0.61$, thus, the flow rate has a maximum. While the existence of this maximum is predicted by the theory, we cannot make a direct comparison since the method to calculate the friction coefficient fails at high concentrations. 

To highlight the effect of the slip we write the flow rate difference between the slip and the no-slip situations using the  
last term in Eq. (\ref{eq:Qapprox}) 
\begin{equation}
\frac{4hq\rho_0\lambda^2E_x}{\eta_0} \ \frac{L_s \tan(h/\lambda)}{\lambda} = 
\frac{4 h E_x}{\eta_0} |\Sigma_\text{wall}|L_s \, ,
\end{equation}
where $|\Sigma_{\textrm{wall}}|$ represents the absolute value of the surface charge density. If $Q_{L_s=0}$ is the flow rate for zero slip, we can define the theoretically predicted enhancement coefficient \cite{hansen_2022}
\begin{equation}
E = \frac{Q}{Q_{L_s =0 }} = 1 + \frac{4 h E_x}{\eta_0 Q_{L_s=0}} |\Sigma_\text{wall}|L_s .
\label{eq:enhancement}
\end{equation}
This enhancement coefficient is plotted in the inset of Fig. \ref{fig:flowratecmp}. Notice that the product $|\Sigma_\text{wall}|L_s$ determines the flow enhancement, and that the enhancement decreases as a function of concentration due to the increasing friction coefficient (or equivalent to the decreasing slip length). 

\begin{table}
\caption{Summary of the system properties with ion charge $q=+1.6$. Symbols are the same as in Table \ref{tab:sys_info_q02} and again $h = $ 9.7. \label{tab:sys_info_q16}}
\begin{ruledtabular}
    \begin{tabular}{cccccc}
    \noalign{\vskip 2mm} 
    $x$  & $\rho_0$ & $\lambda$ & $\xi_0$ & $\Sigma_\text{wall}$ & $Q$ (NEMD)\\
    \noalign{\vskip 2mm} 
    \hline 
    \noalign{\vskip 2mm} 
    0.014    & 0.0016 & 6.59 &  1.46 $\pm$ 0.07 & -0.2 & 3.8 $\pm$ 0.2\\
    0.028    & 0.0017 & 6.42 &  1.81 $\pm$ 0.05 & -0.4 & 5.6 $\pm$ 0.2 \\
    0.042    & 0.0018 & 6.36 &  2.72 $\pm$ 0.10 & -0.6 & 5.3 $\pm$ 0.2 \\
    0.049    & 0.0016 & 6.33 &  3.44 $\pm$ 0.10 & -0.7 & 5.1 $\pm$ 0.2 \\
    0.062    & 0.0015 & 6.31 &  5.87 $\pm$ 0.20 & -0.9 & 3.9 $\pm$ 0.2\\
    \noalign{\vskip 2mm} 
    \end{tabular}
    \end{ruledtabular}
\end{table}

\begin{figure}
    \includegraphics[width=0.45\textwidth]{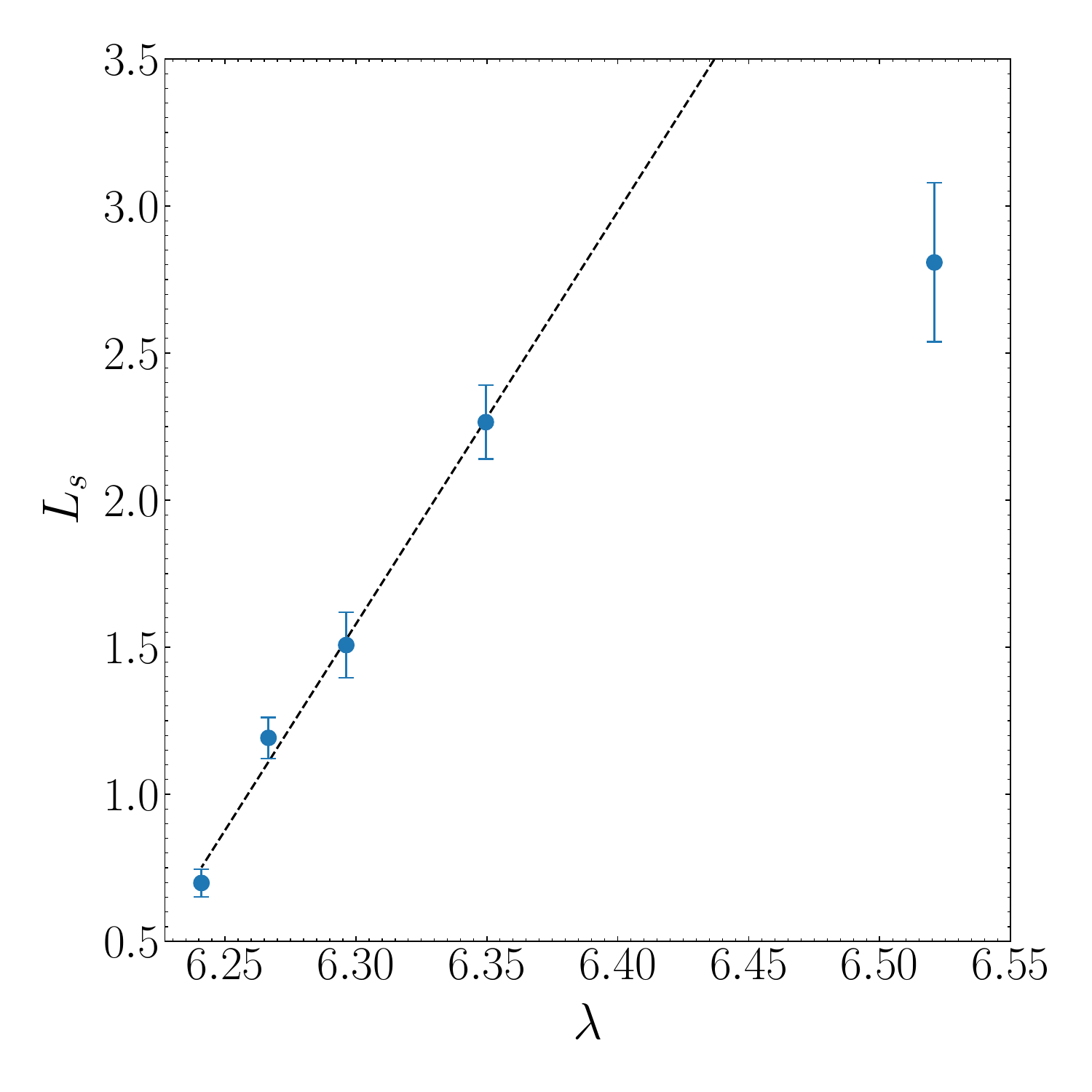}
    \caption{Slip length, $L_s$, versus screening length, $\lambda$, for $q=+1.6$. Circles
    with error bars are results derived from EMD simulation data, and the dashed line is a linear fit of the lowest 4 data points. 
    }
    \label{fig:lsvslambda}
\end{figure}

We now switch to the case where the ion charge is $q = +1.6$; the system properties are listed in Table \ref{tab:sys_info_q16}. 
As it is seen directly from the table values, this system features a maximum flow rate in the regime where we can calculate the friction coefficient. To compare the prediction of the maximum with simulation data, Fig. \ref{fig:lsvslambda} plots the correlation between the slip length $L_s$ and the screening length $\lambda$. For low $\lambda$, corresponding to large ion concentrations, $L_s$ increases linearly with $\lambda$ indicating in this regime the wall-fluid interactions (and resulting fluid layering near the wall) are affected by the presence of the ions. As $\lambda$ increases (the concentration decreases) we can expect the wall-fluid interactions to be dominated by the solvent particle layering; this picture agrees with the observation that the slip length becomes less dependent on the screening length in the 
low concentration regime. 

According to the theory, a maximum flow rate may exist if the slip length is correlated with the screening length. To investigate this, we perform a linear fit to the lowest four data points giving a correlation function $L_s(\lambda) = 14.03\lambda  - 86.79$, see Fig. \ref{fig:lsvslambda}. Substitution into Eq. (\ref{eq:lambda0}) we obtain a maximum flow rate at $\lambda_\text{max} = 6.4$, or equivalently, at concentration $x=0.04$. This is in good agreement with the data in the table. 

\begin{figure}
\centering
\includegraphics[width=0.45\textwidth]{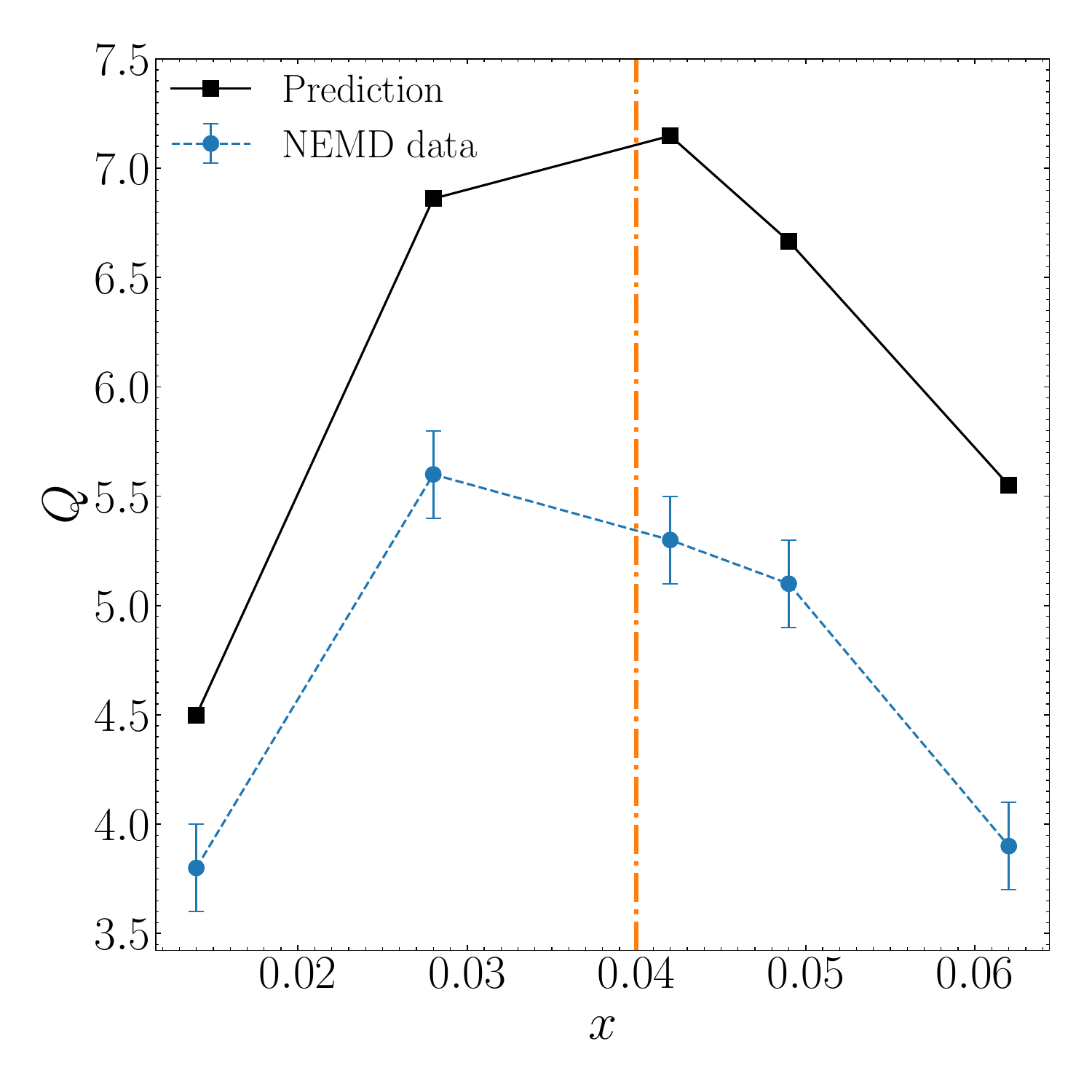} 
\caption{
Volumetric flow rate versus counterion concentration for $q = +1.6$. Circles connected with the dashed line are the results from the NEMD simulations and squares connected with the solid line are the theoretical predictions. The vertical dash-dot line indicates 
the predicted counterion concentration giving the maximum electro-osmotic flow rate.}
\label{fig:flowratecmp_q_1_6}
\end{figure}

In Fig. \ref{fig:flowratecmp_q_1_6} we compare the predicted flow rate and the flow rate obtained from NEMD simulations for $q = +1.6$. While the predictions are in qualitative agreement with the simulation data, the predicted flow rate is overestimated for the systems studied here. If we compare with the case for $q=+0.2$ and at the corresponding wall charges which is a measure for the overall system charge, one can conclude that the theory performs better for lower ion charge and that, in general, the EOF properties depend on both charge density and ion charge.

\begin{figure}
\centering
\includegraphics[width=0.45\textwidth]{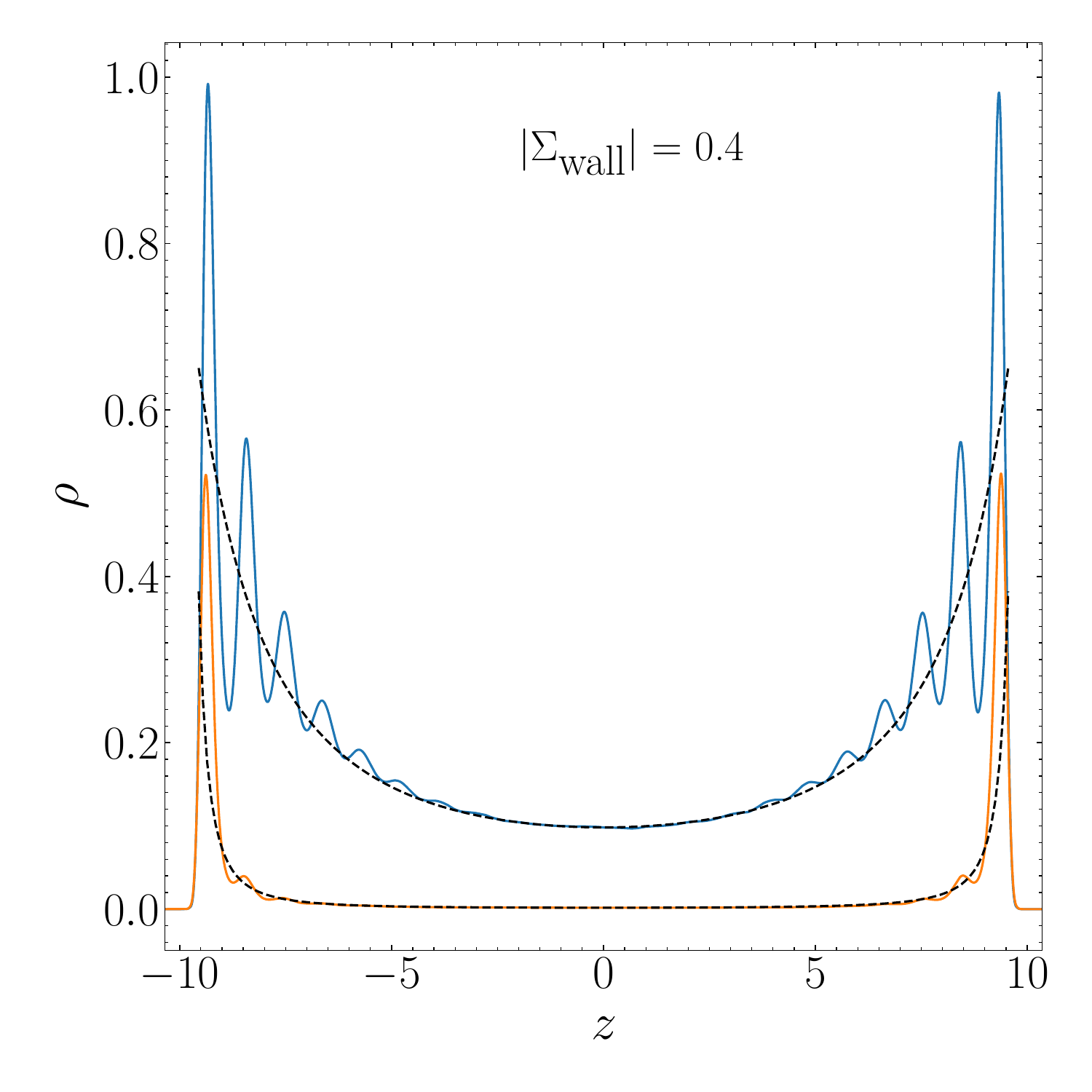} 
\caption{
    Density profiles for $q=+0.2$ and $q=+1.6$. Both systems have the same overall charge density $|\Sigma_\text{wall}|=0.4$. Full lines are simulation results and 
    black dotted lines are the PB equation prediction.  
    \label{fig:chargedenscmp}
}
\end{figure}

The reason for the latter effect is due to the different ion layering at the wall-fluid surface. Figure \ref{fig:chargedenscmp} shows the ion density profiles for $q=+0.2$ and $q=+1.6$ at same system charge $|\Sigma_\text{wall}| = 0.4$. For $q=+0.2$ this corresponds to the concentration $x=0.22$, where the velocity predictions are excellent and the bulk ion number density profile follows the PB equation. For $q=+1.6$ the concentration is $x=0.028$, that is, an order of magnitude lower than for $q=+0.2$ with the same system charge. It can be seen that almost all ions are located in the wall-fluid interface and that the screening length is reduced dramatically compared to the case of $q=+0.2$. For small screening length the driving force acts primarily in the wall-fluid interface where the PB equation does not capture the correct distribution and the theory predictions will deviate significantly from the simulation data. 

\section{Conclusions}
In summary, we have devised an electro-hydrodynamic description for a counterion-only flow system that correctly incorporates fluid slip and therefore enables a correct prediction of the EOF for sufficiently small charge densities and ion charges. The inclusion of slip is based on the independent calculation of the friction coefficient parameter and the method relies only on the interface particles and we can attribute the friction coefficient to be an intrinsic property of the solid-fluid interface. 

The main result of this study is that the EOF features a maximum volumetric flow rate. This is due to the correlation between the slip length and the screening length; the latter itself being dependent on the system ion concentration. In no-slip systems, the flow is monotonically increasing with ion concentration and does not feature this non-trivial maximum.  The result was predicted by the continuum theory and confirmed through direct NEMD simulations. This also implies that the continuum theory correctly predicts the flow for sufficiently low charge density, even if the charge density profile does not follow the simple PB predictions. Finally, we also find that the flow properties depend on the charge of the counterion present in the solution. 
  
The result implies that it is possible to tune the electro-osmotic flow in a counterion-only system if we have accurate values for the hydrodynamic slip. A specific application of this could be in improving the efficiency of water flow through nanochannels coated with polyelectrolytes; recent studies on surfaces coated with polyelectrolytes have reported the possibility of maximizing the EOF  by controlling the nature of the confining substrate \cite{barker2000control, senechal2017electrowetting}. In contrast to the present study, for aqueous solutions, it is important to consider the anisotropic dielectric nature of water under nanoconfinement \cite{bonthuis2013beyond, bonthuis2012profile, varghese2019effect}. Hence, to optimize the flow in these aqueous solutions we will need to understand the influence of dielectric anisotropy on the electro-osmotic behavior of confined systems.

\begin{acknowledgments}
This work was performed on the OzSTAR national facility at Swinburne University of Technology. 
The OzSTAR program receives funding in part from the Astronomy National Collaborative Research Infrastructure Strategy (NCRIS) allocation provided by the Australian Government, and from the Victorian Higher Education State Investment Fund (VHESIF) provided by the Victorian Government.  The authors wish to thank Professor Peter Daivis for the helpful discussions of our work.
\end{acknowledgments}

\appendix
\section{Supporting Information}
The supporting information contains further theoretical details and results relevant to this work.
It includes details on the estimation of viscosity and a detailed theoretical description of the friction coefficient methodology employed in this work.

\bibliography{aiptemplate}

\end{document}


\preprint{AIP/123-QED}

\title{Supplementary Information: Existence of a maximum flow rate in electro-osmotic systems} 

\author{Sleeba Varghese}
\affiliation{ 
Sorbonne Université, CNRS, Physico-chimie des Électrolytes et Nanosystèmes Interfaciaux, PHENIX, F-75005 Paris}%
\author{Billy D. Todd}
\affiliation{Department of Mathematics, School of Science, Computing and Engineering Technologies, Swinburne University of Technology, Melbourne, Victoria 3122, Australia}
\author{Jesper.S. Hansen}
\email{jschmidt@ruc.dk}
\affiliation{``Glass and Time'', IMFUFA,
Department of Science and Environment, Roskilde University, Roskilde 4000, Denmark}

\date{\today}

\pacs{}

\maketitle 

\section{Friction Coefficient Methodology}
The friction coefficient is calculated from the model proposed by 
Hansen et al.\cite{hansen2011prediction}, where the value for interfacial friction coefficient is obtained from the statistical fluctuations of a slab of particles close to the wall.
For sufficiently small relative velocities, the linear constitutive equation that relates the wall-slab shear force to the relative velocity of the slab is  given by,
\begin{eqnarray}
F_{x}(t) = -\int_{0}^{t}\zeta(t-\tau)\Delta u_{x}(\tau)d\tau + F_{r}(t), \label{eq:const_eq}
\end{eqnarray}
where $F_{x}$ is the wall-slab shear force, $\zeta$ is the friction kernel and $\Delta u_{x} (=u_{slab,x}-u_{wall,x})$ is the $x$-component of relative velocity between wall and slab.
For immobile walls, $\Delta u_{x} = u_{slab,x} = u_x$.
$F_r$ is a random force term that has zero mean and is uncorrelated with $u_{slab,x}$.
Multiplying Eq. (\ref{eq:const_eq}) with $u_{slab,x}$ and 
taking the ensemble average will provide us with a corresponding relation between the slab velocity-force cross-correlation function $C_{u_{x}F_{x}}(t)$, and the slab velocity autocorrelation function $C_{u_{x}u_{x}}(t)$,
\begin{eqnarray}
C_{u_{x}F_{x}}(t) = -\int_{0}^{t} \zeta(t-\tau)C_{u_{x}u_{x}}(\tau)d\tau, \label{eq:integ_eq}
\end{eqnarray}
where $C_{u_{x}F_{x}}(t) = \big< u_x(0)F_x(t)\big>$ and
$C_{u_{x}u_{x}}(t) = \big< u_x(0)u_x(t)\big>$.
The Laplace transform of Eq. (\ref{eq:integ_eq}) converts the integral expression 
into a more convenient algebraic expression
\begin{eqnarray}
{{\tilde C}_{{u_x}{F_{x}}}}(s) = 
-{\tilde \zeta}(s){{\tilde C}_{{u_x}{u_x}}}(s), \label{eq:interg_eq_LT}
\end{eqnarray}
where $s$ is the Laplace coordinate and the Laplace transform of a 
function $f(t)$ is defined as
\begin{eqnarray}
L[f(t)] = \int_{0}^{\infty} f(t) e^{-st} dt = \tilde{f}(s). \label{eq:LT_def}
\end{eqnarray}

According to Ref.~\citenum{hansen2011prediction}, the friction kernel $\zeta$ can be approximated as a one-term Maxwellian memory function, i.e.,
\begin{eqnarray}
\zeta(t) = B_{1} e^{\lambda_{1}t}. \label{eq:memory_function_def}
\end{eqnarray}
which implies
\begin{eqnarray}
{\tilde \zeta}(s) = \frac{B_{1}}{s+\lambda_{1}}, \label{eq:fric_kernel}
\end{eqnarray}
where $B_{1}$ and $\lambda_{1}$ are the Maxwellian parameters.
Substituting Eq. (\ref{eq:fric_kernel}) in (\ref{eq:interg_eq_LT}) we get
\begin{eqnarray}
{{\tilde C}_{{u_x}{F_{x}}}}(s) = 
-\frac{B_{1}}{s+{\lambda_{1}}} {{\tilde C}_{{u_x}{u_x}}}(s). \label{eq:corr_exp}
\end{eqnarray}
Now to obtain the values for the Maxwellian parameters $B_{1}$ and $\lambda_{1}$ we follow the methodology suggested by Varghese et al.\cite{varghese2021improved}, where 
Eq. (\ref{eq:corr_exp}) is rearranged and integrated from an arbitrary minimum $\beta_i$ to an arbitrary maximum $\beta_j$
\begin{eqnarray}
\int_{\beta_i}^{\beta_j} \frac{s+{\lambda_{1}}}{{B_{1}}} ds &=& 
-\int_{\beta_i}^{\beta_j}\frac{{{\tilde C}_{{u_x}{u_x}}}(s)}
{{{\tilde C}_{{u_x}{F^{'}_{x}}}}(s)} ds
\hfill \nonumber \\
\implies \frac{1}{{B_1}}\bigg[\frac{\beta_{j}^2 - \beta_{i}^2}{2} + {\lambda_1}(\beta_j - \beta_i) \bigg] &=& \Lambda, \label{eq:varghese_method}
\end{eqnarray}
where $\Lambda = -\int_{\beta_i}^{\beta_j}\frac{{{\tilde C}_{{u_x}{u_x}}}(s)}
{{{\tilde C}_{{u_x}{F^{'}_{x}}}}(s)} ds$, which can be numerically determined 
directly from equilibrium molecular dynamics simulations. 

To solve for the parameters $B_1$ and $\lambda_1$ we need two equations, which are obtained by taking two different integral limits for Eq. (\ref{eq:varghese_method}), i.e., $\beta_1$ to $\beta_2$ and $\beta_3$ to $\beta_4$,
\begin{eqnarray}
\frac{1}{{B_1}}\bigg[\frac{\beta_{2}^2 - \beta_{1}^2}{2} + {\lambda_1}(\beta_{2} - \beta_{1}) \bigg] &=& \Lambda_{1}
\label{eq:integ_limit_1}
\end{eqnarray}
and
\begin{eqnarray}
\frac{1}{{B_1}}\bigg[\frac{\beta_{4}^2 - \beta_{3}^2}{2} + {\lambda_1}(\beta_{4} - \beta_{3}) \bigg] &=& \Lambda_{2}.
\label{eq:integ_limit_2}
\end{eqnarray}
Dividing Eq. (\ref{eq:integ_limit_1}) by Eq. (\ref{eq:integ_limit_2}) and
taking $\Lambda_1/\Lambda_2 = C$ we find
\begin{eqnarray}
\bigg[\frac{\beta_{2}^2 - \beta_{1}^2}{2} + {\lambda_1}(\beta_{2} - \beta_{1}) \bigg] &=&
C\bigg[\frac{\beta_{4}^2 - \beta_{3}^2}{2} + {\lambda_1}(\beta_{4}-\beta_{3}) \bigg].
\label{eq:integ_limit_final_exp}
\end{eqnarray}
Rearranging Eq. (\ref{eq:integ_limit_final_exp}) we get
\begin{eqnarray}
\lambda_1 &=& \frac{{\Big(\beta_{2}^2 - \beta_{1}^2} - {C(\beta_{4}^2 - \beta_{3}^2)}\Big)}{2\Big(C(\beta_4 - \beta_3) - \beta_2 + \beta_1 \Big)}. \label{eq:lamda1}
\end{eqnarray}
By substituting the value of ${\lambda_1}$ in Eq. (\ref{eq:integ_limit_1}) or
(\ref{eq:integ_limit_2}) we can obtain the value for ${B_1}$.
Finally, the steady-state fluid-solid intrinsic interfacial friction coefficient $\xi_0$
is obtained as
\begin{eqnarray}
\xi_{0} = \frac{\tilde{\zeta}(0)}{A} = \frac{B_1}{A\lambda_1}, \label{eq:friction_coeff}
\end{eqnarray}
where $A$ is the wall surface area.

\begin{figure}
\begin{tabular}{cc}
\includegraphics[width=0.5\textwidth]{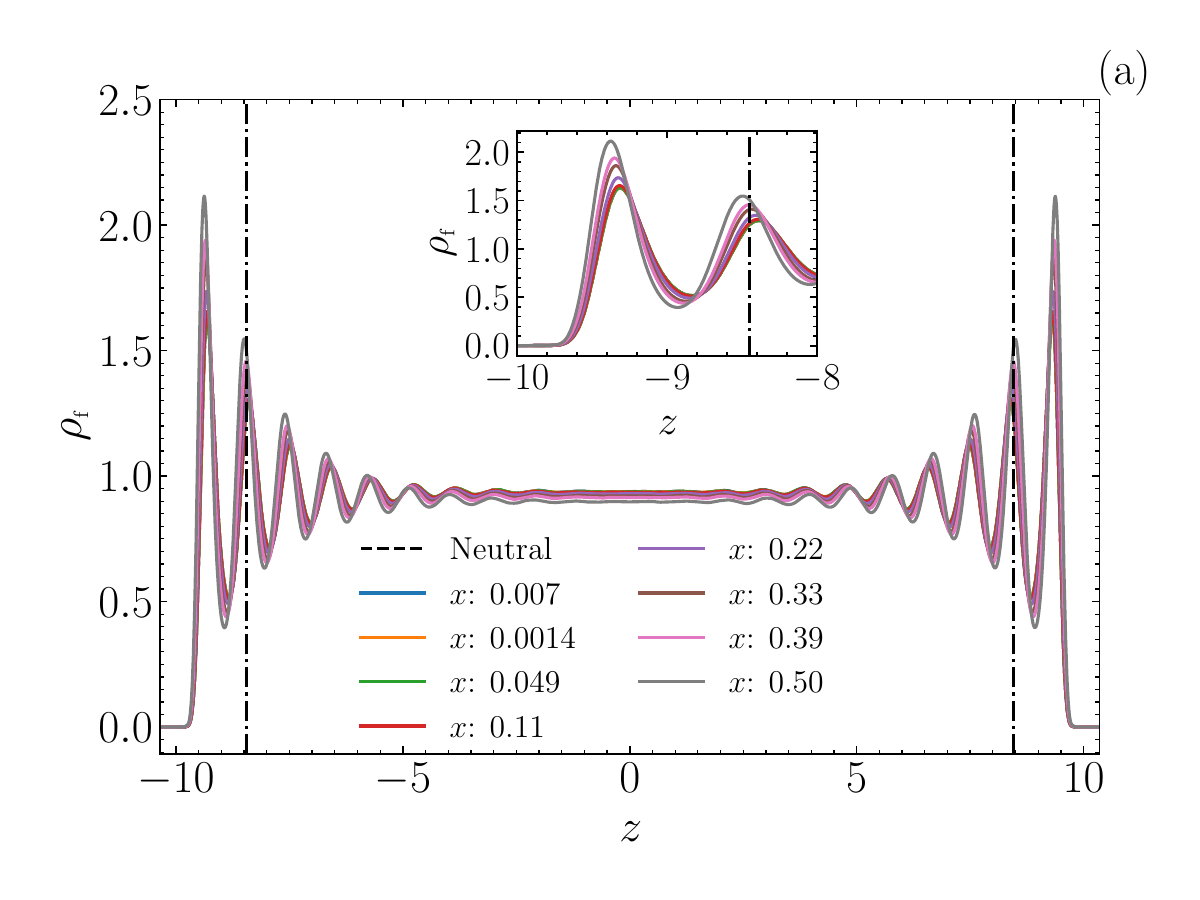} &
\includegraphics[width=0.5\textwidth]{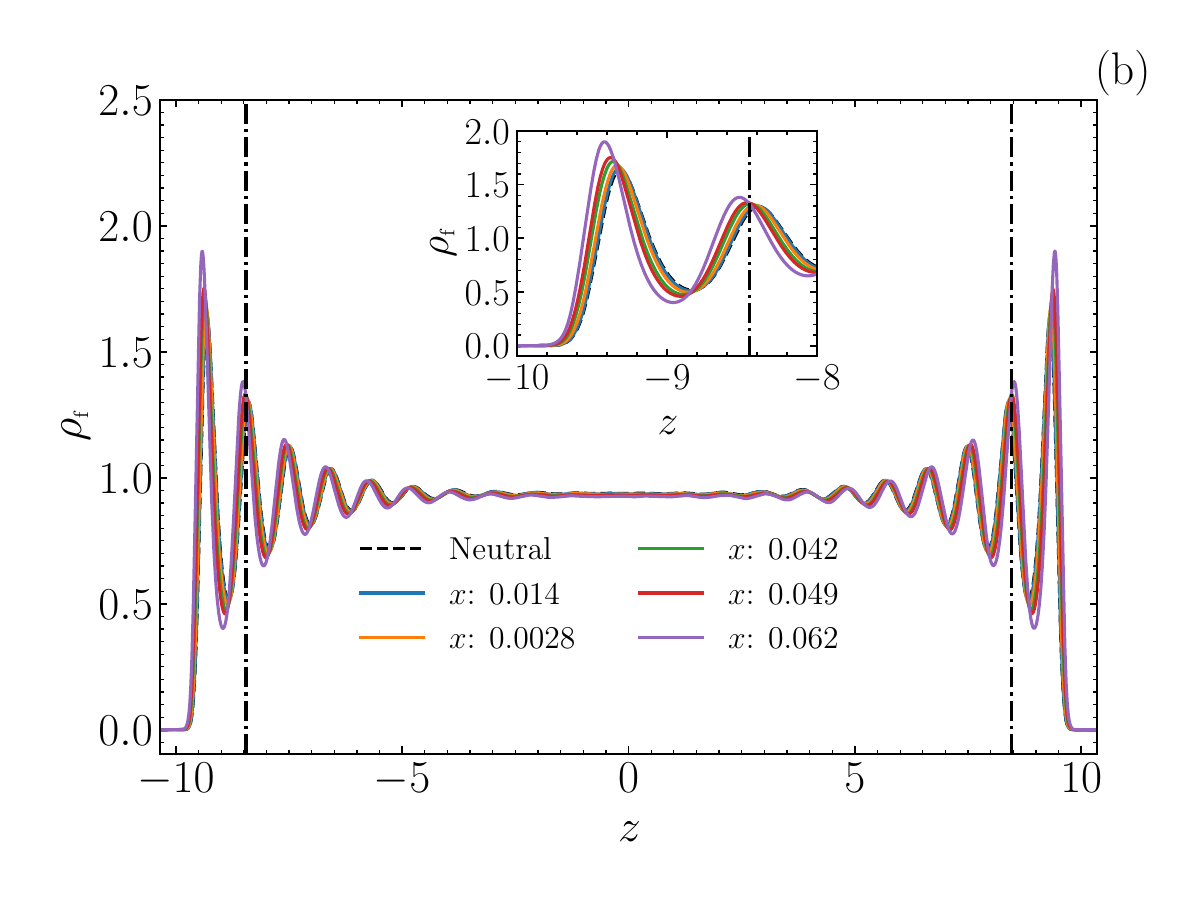} \\
\end{tabular}
\caption{Fluid density profile, $\rho_{\textrm{f}}$, for different counterion concentrations for counterion charge, (a) $q = +0.2$, and (b) $q = +1.6$.
The inset plot shows a magnified version of the slab region adjacent to the bottom wall.
The black dash-dot vertical lines (in the main and inset) indicate the slab width coordinates from the walls, and the neutral system corresponds to the system without any counterions.
}
\label{fig:fluid_density}
\end{figure}

One important parameter to consider in estimating the interface intrinsic friction coefficient is the slab width used in its computation.
An optimal value for slab width should have a sufficient amount of interfacial particles to provide a statistically good approximation for friction coefficient but should not be large enough to include effects due to viscous forces.
Hence in this study, we consider a slab of particles $2\sigma$ away from each wall to compute the interfacial friction coefficient.
The slab width of $2\sigma$ was chosen in order to encompass the
first fluid density layer (Figures \ref{fig:fluid_density} (a) and (b)), which is shown to have the crucial role in determining the value of the friction coefficient.\cite{niavarani2008slip}

\section{Viscosity}
\begin{figure}
\begin{tabular}{cc}
\includegraphics[width=0.5\textwidth]{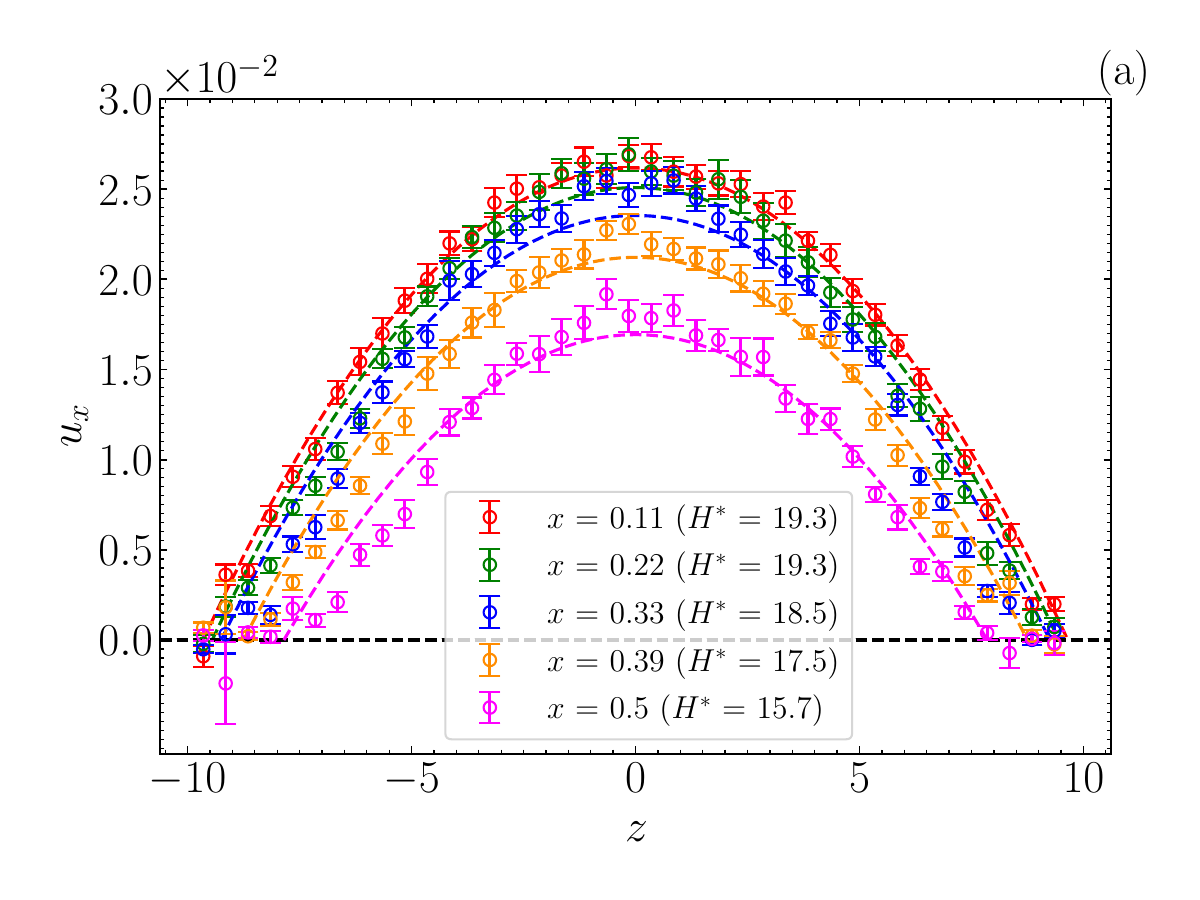} &
\includegraphics[width=0.5\textwidth]{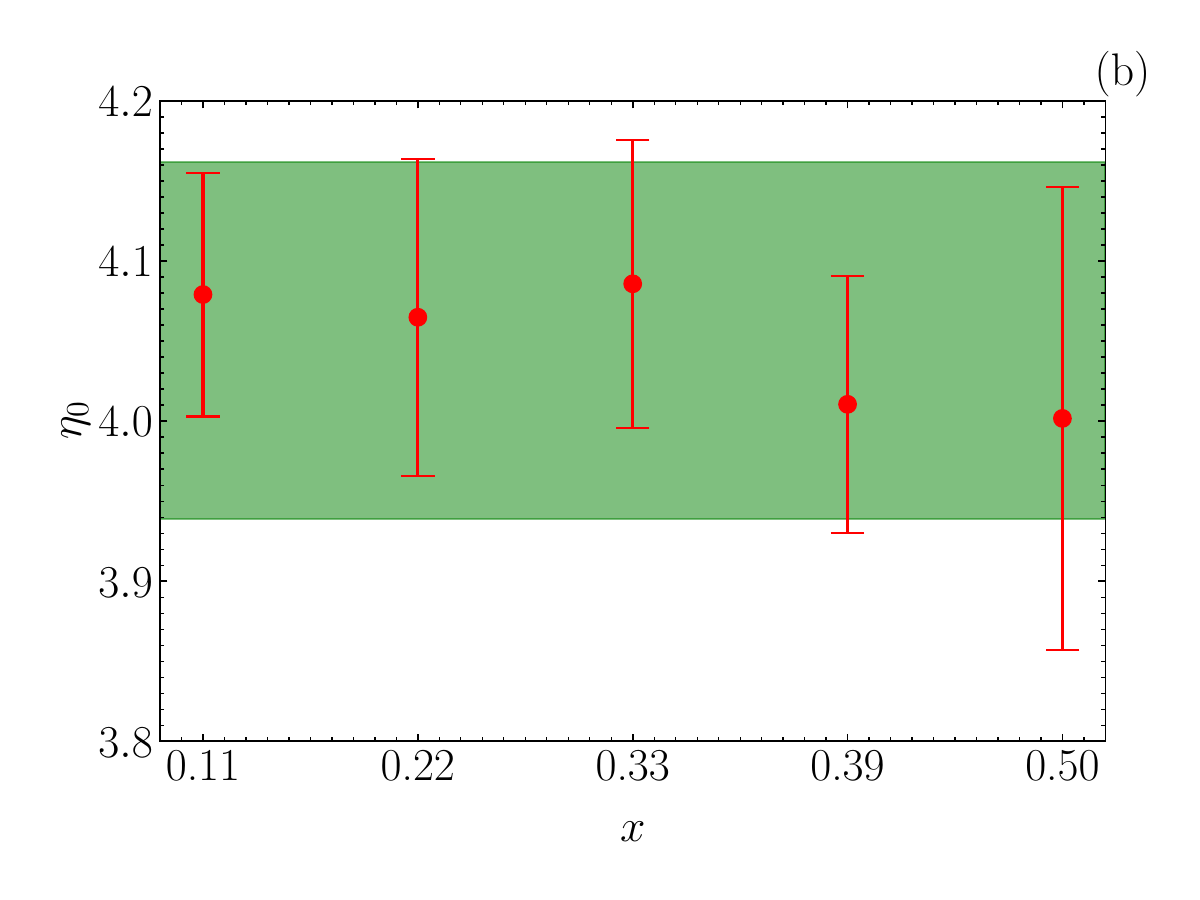} \\
\end{tabular}
\caption{(a) Poiseuille flow velocity profiles, $u_{x}$, at different counterion concentrations under an external field, 
$F_e = 0.0025$. 
The markers denote the NEMD velocity profile, and the dashed lines represent the curve fit.
(b)Variation of viscosity with counterion concentration. 
The shaded region represents the viscosity range for a neutral or uncharged system.
Each value of viscosity was averaged over 20 independent NEMD simulations.
Literature value for the bulk LJ system at $\rho = 0.9$,
and $T = 1.0$ is $4.060 \pm 0.827$ \cite{rowley1997diffusion}.}
\label{fig:visc}
\end{figure}

To compute the effective shear viscosity, we perform non-equilibrium molecular dynamics (NEMD) simulations with an external acceleration field applied along the $x$-direction to generate the Poiseuille flow inside a wetting nanochannel.
The Poiseuille flow solution for the velocity profile with confinement along the $z$-direction is given as,
\begin{eqnarray}
u_x(z) &=& \frac{\rho^{\textrm{bulk}}_{\textrm{f}} F_e}{2\eta_0}\Bigg[\Big(\frac{H^{*}}{2}\Big)^{2} - z^{2}\Bigg], \label{eq:poise_equ}
\end{eqnarray}
where $\rho^{\textrm{bulk}}_{\textrm{f}}$ is the bulk average density of the confined fluid, $F_e$ is the external field applied, and $H^{*}$ denotes the channel width available for the flow, which is taken as the distance between the $z$-coordinate values where $u_{x}(z)$ reaches zero.
The effective viscosity $\eta_{0}$ for a particular counterion system is calculated by curve fitting its NEMD velocity profile with Eq. (\ref{eq:poise_equ}) (Figure \ref{fig:visc}(a)).

The wetting nanochannel is modeled by choosing a solid-fluid wetting 
coefficient $C_{\textrm{FS}} = 1.0$, which is shown to model a hydrophilic surface 
at a reduced temperature, $T = 1.0$ \cite{barrat1999large}.
Figure \ref{fig:visc}(b) shows the variation of viscosity with counterion concentration,
for systems with counterion charge $q = +0.2$.
Since the viscosity values for counterion systems fall within the acceptable viscosity range of a neutral system, we report here that the 
effective shear viscosity does not vary significantly with the counterion concentration range
investigated in this study.
Hence for counterion concentrations below 0.11, we use the effective shear viscosity value of a
neutral system for the electro-osmotic flow calculations.

\bibliography{suppinfo}